\documentclass[journal]{IEEEtran}
\usepackage{cite}
\ifCLASSINFOpdf
\else
\fi
\usepackage{amsmath}
\usepackage{algorithmic}
\usepackage{url}

\usepackage{epsfig}
\usepackage{amsthm} % for split/cases use
\usepackage{mathrsfs}   % for /mathca use
\usepackage{amsfonts} % for non-negative integer symbol
\usepackage{multirow} % for multi-rows table
\usepackage{graphicx}
\usepackage{subfigure}
\usepackage{algorithm}
\usepackage{algorithmic}
\usepackage{dcolumn}
\usepackage{booktabs}
\usepackage{mathtools}
\usepackage[usenames, dvipsnames]{color}
\definecolor{myblue}{rgb}{0.0, 0.0, 1}

\usepackage{dcolumn}
\newcolumntype{d}{D{.}{.}{2.2}}  %2.2 是调节数字的居中位置的，可以更改

\begin{document}

%\title{Bare Demo of IEEEtran.cls\\ for IEEE Journals}
%\title{A Two-phase On-line Joint Scheduling of Pricing and Charging Control for Charging Station Considering Price Fluctuation Sensitivity}
\title{A Two-phase On-line Joint Scheduling for Welfare Maximization of Charging Station}

%\author{Michael~Shell,~\IEEEmembership{Member,~IEEE,}
%        John~Doe,~\IEEEmembership{Fellow,~OSA,}
%        and~Jane~Doe,~\IEEEmembership{Life~Fellow,~IEEE}% <-this % stops a space
%\thanks{M. Shell was with the Department
%of Electrical and Computer Engineering, Georgia Institute of Technology, Atlanta,

\author{Qilong~Huang,~\IEEEmembership{Member,~IEEE},
        Qing-Shan~Jia,~\IEEEmembership{Senior~Member,~IEEE},
        Xiang~Wu,~\IEEEmembership{Member,~IEEE},
        Shengyuan~Xu,
        and~Xiaohong~Guan,~\IEEEmembership{Life Fellow,~IEEE}
        % <-this % stops a space
%\thanks{{\color{myblue}{This work is supported in part by NSFC (61222302, 61174072, 61221063, 91224008, U1301254, and 61425027), the Tsinghua-Leuven Collaboration Project, 111 International Collaboration Program of China (B06002), the National Key Technology R\&D Program (2013BAG18B00), TNList Funding for Cross Disciplinary Research, and the Program of New Star in Science and Technology in Beijing (No. xx2014B056).}}}% <-this % stops a space
\thanks{{This work has been submitted to the IEEE for possible publication. Copyright may be transferred without notice, after which this version may no longer be accessible.}}
%\thanks{{This work was supported by the National Natural Science Foundation of China under grants No. 62103191, No. 62125304, No. 62073182, No. 62192751 and the Natural Science Foundation of Jiangsu Province (Grant No. BK20210336).}}
\thanks{Q. Huang, X. Wu and S. Xu are with the School of Automation, Nanjing University of Science and Technology, Nanjing, 210094, China (email: huangql@njust.edu.cn, wuxiang1@njust.edu.cn, syxu@njust.edu.cn).}% <-this % stops a space
\thanks{Q.-S. Jia and X. Guan are with CFINS, Department of Automation, Tsinghua University, Beijing, 100084, China (email: jiaqs@tsinghua.edu.cn, xhguan@tsinghua.edu.cn).}% <-this % stops a space
\thanks{X. Guan is also with MOE KLINNS Lab, Xi'an Jiaotong University, Xi'an, China, 710049.}
%\thanks{Li. Yang is the corresponding author.}% <-this % stops a space
}

\maketitle

\begin{abstract}

The large adoption of EVs brings practical interest to the operation optimization of the charging station. The joint scheduling of pricing and charging control will achieve a win-win situation both for the charging station and EV drivers, thus enhancing the operational capability of the station. We consider this important problem in this paper and make the following contributions. First, a joint scheduling model of pricing and charging control is developed to maximize the expected social welfare of the charging station considering the Quality of Service and the price fluctuation sensitivity of EV drivers. It is formulated as a Markov decision process with variance criterion to capture uncertainties during operation. Second, a two-phase on-line policy learning algorithm is proposed to solve this joint scheduling problem. In the first phase, it implements event-based policy iteration to find the optimal pricing scheme, while in the second phase, it implements scenario-based model predictive control for smart charging under the updated pricing scheme. Third, by leveraging the performance difference theory, the optimality of the proposed algorithm is theoretically analyzed. Numerical experiments for a charging station with distributed generation and energy storage demonstrate the effectiveness of the proposed method and the improved social welfare of the charging station.\\

%\textit{Note to Practitioners---}
%The popularization of EVs requires the high-efficiency operation of the charging station. The joint scheduling of pricing and charging control can provide a promising way to achieve balance between EV drivers' satisfaction and the profit maximization of the charging station. However, it suffers from the uncertain responses of EV drivers to the pricing scheme and the coupled relationship between pricing and charging control. This multi-stage stochastic programming is non-trivial to solve. In this paper, we propose a two-phase on-line policy learning method for this joint scheduling problem. This algorithm can be implemented in the controller of the charging station. In the first phase, the event-based policy iteration can be implemented to iteratively improve the current best pricing scheme until convergence. In the second phase, the scenario-based model predictive control which reformulated as a traditional mixed integer linear programming can be quickly solved for smart charging. Case studies show the operation enhancement of the charging station.
\end{abstract}


% \fi
\IEEEpeerreviewmaketitle

\section{Introduction}
% form to use if you need the single drop letter followed bIEEEhowto:kopkay
\IEEEPARstart{A}{cting} as the main hinge between transportation sector and power sector, electric vehicles (EVs) attract more and more attention in recent years. On one hand, the EVs will largely reduce the carbon emission of the transportation sector if they are supplied by clean energy, such as wind power, solar power, etc. On the other hand, the EVs can be used as mobile storage to increase the demand elasticity in the power sector. Due to these reasons, many countries have made incentive policy to stimulate the EV market. For example, as the world's largest EV market, several cities in China, such as Beijing, Shenzhen, are subsidizing to accelerate the shift to electric taxi and it is estimated that there will be over 300 thousand electric taxies until 2020\cite{WebsiteEVPolicy}.

The main influential factors for the popularization of EVs lie in two aspects. The first is the charging price which is the main focus of the EV drivers, especially for electric taxi drivers. The current unsustainable low price for EV charging comes from various incentive subsidies comparing with relatively expensive oil price \cite{li2011pricing}. Another is the charging control which is the main focus of the charging station. Existing literatures have shown that disorderly charging will incur increased operation cost of the charging station and may bring in damages to the power grid \cite{das2020electric}. Luckily, these factors can also be controlled by the charging station. Therefore, it is of great practical interest for the charging station to optimize the charging price and implement smart charging control for the EVs in order to achieve a win-win situation for both the charging station and EV drivers.

However, this problem usually faces the following challenges. \textit{First}, the uncertainties in the charging behavior of EV drivers. The charging price scheme will influence the EV driver's decision to enter for charging and each driver has different response to the charging price scheme. Furthermore, the arrival time, required charging energy and parking time of EVs are all uncertain before parking to charge. \textit{Second}, the tradeoff between maximizing operation profit and minimizing the charging cost. The low charging price will attract more EVs to enter for charging, however the operation profit may be reduced for the charging station. On the contrary, the high charging price may gain the high unit profit per EV, but the total operation profit may be low as the service number will be reduced due to high charging price. \textit{Third}, the price fluctuation sensitivity of EV drivers. The EV drivers are serious and sensitive to the price fluctuation. Although the large price fluctuation of the charging station may increase the operation profit, it will also bring in uncertainties and variation of charging cost for the EV drivers. This may incur reduced preference to choose this charging station for EV drivers. \textit{Fourth}, the multi-stage coupled relationship between pricing and charging control. The current charging price will influence the number of EVs to be charged and further influence the charging control in the future considering the relative long charging time. The pricing and charging control should be jointly considered in a multi-stage decision fashion.

Based on the discussions above, we study the joint scheduling of pricing and charging control for the charging station in this paper. Compared with the published literature, the main contributions of this paper are as follows:

\begin{itemize}
\item[$\bullet$] A joint scheduling model of pricing and charging control is developed to maximize the expected social welfare of the charging station considering the Quality of Service (QoS), the price fluctuation sensitivity of drivers and uncertain charging behaviors. This model is formulated as a Markov decision process (MDP) with variance criterion to capture uncertainties during operation.
\item[$\bullet$] A two-phase online policy learning method is proposed to solve this joint scheduling problem. In the first phase, an event-based policy iteration which alleviates the burden of large state/action space is developed to find the optimal pricing scheme. In the second phase, the scenario-based model predictive control (MPC) is developed to achieve smart charging under the updated pricing scheme.
\item[$\bullet$] The optimality of the proposed method is theoretically analyzed by leveraging performance difference theory. Numerical results for a typical charging station with distributed generation and energy storage demonstrate the effectiveness of the proposed optimization method and the improved social welfare of the charging station.
\end{itemize}

The rest of this paper is organized as follows. We briefly review the related literature in Section II, formulate the problem in Section III, present the solution methodology in Section IV, discuss the numerical results in Section V, and briefly conclude in Section VI.

\section{Literature Review}
The charging control of EVs has attracted a lot of attention in recent years. In most of works, the charging price is assumed to be known and uncontrollable\cite{9326417,jin2020joint,9088147}. In this case, various charging control methods are proposed for charging process optimization, such as mixed-integer programming\cite{koufakis2019offline}, model predictive control\cite{yang2018decentralized}, reinforcement learning\cite{9163332}, etc. However, besides the charging process optimization, the pricing is another important control mechanism for the charging station to maximum its operation profit. The EV drivers and charging station will further benefit from the joint scheduling of pricing and charging control.

One of the traditional pricing mechanisms is to use price elastic matrix (PEM) to evaluate the charging demand response\cite{kasani2021optimal,srivatsa2017modeling}. The PEM is defined as the ratio of the change in charging demand to the base charging demand over the change in price to the base price. The optimal pricing scheme can be obtained by evaluating the pricing effect under the PEM criterion. However, the individual response to the price is neglected for each EV driver in this method. The individual response will influence the number of EVs which enter into the charging station and further influence the charging control of EVs.

Another important pricing mechanism is to use queue theory to implement dynamic/static pricing control\cite{xia2018dynamic,zhao2021dynamic}. In \cite{bayram2014pricing}, a M/M/C queue model is proposed to determine the discounted charging price to provide charging service deferral for charging demand shifting. In \cite{zhang2018optimal}, an optimal pricing scheme is designed by formulating the dual-mode charging station as a queue network with multiple servers and heterogeneous service rate. In most of works, the arrival rate of EVs is assumed to be constant. As the drivers are price-sensitive, they may choose other charging stations when observing unsatisfactory charging price. In this way, the arrival rate of EVs will be time-varying.

The game theory is also a natural candidate for the pricing of EVs. In \cite{kapoor2022centralized}, the bi-level pricing scheme is obtained by formulating the competition among EV owners, the aggregator and the distribution system operator as the Stackelberg game. In \cite{qian2021multi}, it considers the price competition problem among multiple charging stations as a game with incomplete information of the market environment. In \cite{kabir2020optimal}, it introduces a game-theoretic algorithm to find a charging rate-dependent pricing mechanism which meets the charging demand at the minimum price. However, few consider the uncertainties in the charging behaviors of EVs. Furthermore, the game-based pricing method requires iterative computations to achieve Nash equilibrium which is time-consuming.

In conclusion, few existing literatures consider the individual response of EV drivers to the price and its impact to the charging control of EVs. Furthermore, to the best of the authors' knowledge, the existing literatures have not considered the price fluctuation sensitivity of the EV drivers which affects the drivers' satisfactory to the charging station. Also, most works do not consider the multi-stage coupled relationship between pricing and charging control with uncertain charging behaviors of EVs. Therefore, we will consider these in this paper.

%Different from our previous works which study the EV charging/discharging control in a microgrid of buildings\cite{huang2022simulation,huang2022event}, this paper will focus on the social welfare maximization of the charging station by joint optimization of pricing and control. Furthermore, a two-phase on-line joint optimization method will be proposed for the station considering the structural property of the problem.

\section{Problem Formulation}

\subsection{System Description}
\begin{figure}
\centering
\setlength{\abovecaptionskip}{-6pt}
\includegraphics[width=0.49\textwidth]{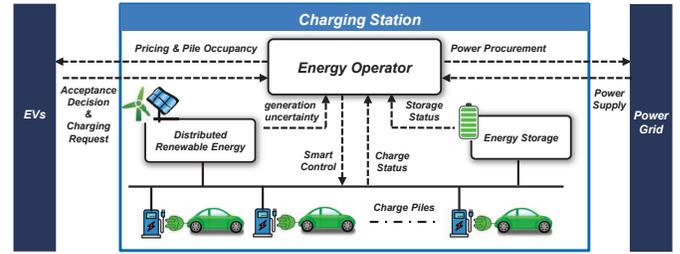}
\caption{System description of the joint scheduling problem.}
\label{SystemOverview}
\end{figure}

We consider a charging station equipped with several charging piles, distributed renewable energy and energy storage as depicted in Fig. \ref{SystemOverview}. The energy operator in the charging station is responsible to determine the pricing scheme and control the charging process of parked EVs. At each decision stage, the energy operator will announce the charging price to EVs based on the distributed energy generation prediction, the status of storage and parked EVs. Then, EVs will determine whether enter into this charging station to charge or not. If accept this charging price, the EVs will enter into the charging station and report their charging requests to the energy operator. Meanwhile, the energy operator should implement smart control for the parked EVs, distributed renewable energy and storage considering the distributed energy generation prediction, the status of storage and parked EVs. The energy operator can also procure electricity from the power grid in case of insufficient supply.

Considering this operation mechanism, the energy operator should jointly implement pricing and charging control of EVs to maximize its social welfare, including the operation profit, QoS and the price fluctuation. As there exists uncertainties in the charging behaviors of EV drivers and the generation of the distributed renewable energy, this problem is a multi-stage stochastic programming problem. In the following, we will use MDP to formulate this problem as it is an effective tool for sequential multi-objective decision-making problem under uncertainty\cite{puterman1990markov}. To simplify the discussions, the fixed charging power is considered to prolong the battery life of EV.

%\IEEEpubidadjcol

\subsection{System Model}
We consider this joint scheduling problem over the discretized horizon $t =1,2,...,T$ where $t$ denotes the decision epoch and $\Delta t$ denotes the decision interval. There are $N$ charging piles in the charging station. The MDP model of this joint scheduling problem is shown below.

\textit{1) System States:} The system state at time $t$ is defined as $s_t=[\overline{p}_t^w,\overline{p}_t^s,b_t,e_t^i,\tau_t^i] \in \mathcal{S}$ where $\overline{p}_t^w$ and $\overline{p}_t^s$ denote the generation of the distributed wind energy and solar power, $b_t$ denotes the State of Charge (SOC) of storage, $e_t^i$ and $\tau_t^i$ denote the remaining required charging energy and remaining parking time of the EV in the $i$th charging pile where $i=1,2,...,N$ and $\mathcal{S}$ denotes the state space. When the $i$th charging pile is unoccupied, there are $e_t^i=0$ and $\tau_t^i=0$.

\textit{2) Actions:} The control action at stage $t$ is defined as $a_t=[\varphi_t,h_t,z_t^1,...,z_t^N] \in \mathcal{A}$ where $\varphi_t$ denotes the charging price, $h_t$ denotes the power output of storage, $z_t^i \in \{0,1\}, i=1,2,...,N$ is the control decision for the $i$th charging pile and $\mathcal{A}$ denotes the action space. $z_t^i = 1$ denotes that the $i$th charging pile will charge its connected EV, otherwise $z_t^i = 0$.

Motivated by \cite{bayram2014pricing}, in order to encourage the EV drivers to join in the smart charging, the charging station will offer a charging price discount for the arrival EVs based on their charging elasticity, i.e., for the $i$th charging pile which has newly arrival EV at stage $t_a$, there is
\begin{equation}
\label{EQPriceDiscount}
\varphi_t^i = \varphi_{t_a} e^{-\theta_p l_{t_a}^i},  \forall t \in [t_a,t_a+\tau_{t_a}^i-1]
\end{equation}
where  $\varphi_t^i$ denotes the charging price for the EV which connected to the $i$th charging pile from $t_a$ to $t_a+\tau_{t_a}^i-1$, $\theta_p$ denotes the discount coefficient, $l_{t_a}^i = \tau_{t_a}^i-e_{t_a}^i/(p\eta^{\text{ev}}\Delta t)$ denotes the charging elasticity of the connected EV, $p$ denotes its charging power and $\eta^{\text{ev}}$ denotes the charging efficiency. It can be seen that the longer parking time and less required charging energy means the larger elasticity. Therefore, the charging price is cheaper. Note that each EV will decide whether enter into the charging station based on its time-invariant discounted price $\varphi_t^i$. When $\varphi_t^i$ is determined, the charging cost will be known by the EV owners by multiplying $\varphi_t^i$ with the known required charging energy.

\textit{3) System Dynamics:} For the charging station, there is the following relationship for the number of parked EVs, i.e.,
\begin{equation}
n_{t+1} = n_t + n_t^{\text{in}} - n_t^{\text{out}}
\end{equation}
where $n_t$ denotes the number of parked EVs at stage $t$, $n_t^{\text{in}}$ denotes the number of newly arrival EVs which decide to enter into the charging station and $n_t^{\text{out}}$ denotes the number of departure EVs due to the end of parking at stage $t$.

Based on the observation, the charging station can obtain the probability distribution of the number of newly arrival EVs. Usually, it can be assumed to follow poisson distribution $f(\lambda_t)$ with mean arrival rate $\lambda_t$ at stage $t$\cite{li2012modeling}. Once arrival, the EV driver will decide whether enter into the charging station based on the announced charging price $\varphi_t$ and the availability of unoccupied charging piles. As EV drivers express difference preferences over the charging price, let $f(\varphi_t)$ denotes the acceptance probability of the charging price. Then, the probability distribution of $n_t^{\text{in}}$ satisfies
\begin{equation}
\label{EQArrival}
n_t^{\text{in}} \sim f(\lambda_t) \cdot f(\varphi_t).
\end{equation}
Based on \cite{jhala2017real}, the Bernoulli distribution can be used to represent the acceptance probability $f(\varphi_t)$, i.e.,
\begin{equation}
\label{EQBernoulli}
f(\varphi_t) \sim \text{Bernoulli}(1-\varphi_t/\overline{\varphi})
\end{equation}
where $\overline{\varphi}$ denotes the upper bound of the charging price. It can be seen that the larger the charging price, the smaller the acceptance probability. Note that the proposed method can also be applied when other representations of acceptance probability are used.

For the EV connected to the $i$th charging pile a stage $t$, there is
\begin{equation}
\label{EQRemainParkTime}
\tau_{t+1}^i = \tau_t^i - 1
\end{equation}
\begin{equation}
e_{t+1}^i = e_t^i - z_t^i p \eta^{\text{ev}} \Delta t
\end{equation}
Note that for the current connected EVs, the values of $\tau_t^i$ and $e_t^i$ are known which will be provided by the drivers. For the future arrival EVs which enter into the charging station at stage $\hat{t}$ where $\hat{t}>t$, the value of $\tau_{\hat{t}}^i$ and $e_{\hat{t}}^i$ can be estimated based on their probability distributions which can be obtained from statistical analysis of the real charging data.

For the generation of distributed renewable energy, based on \cite{sarkar2011mw}, there is
\begin{equation}
\overline{p}_t^w =
\begin{cases}
p^{\text{cap}}_w,& v_{\text{rated}} < v_t \leq v_{\text{cutout}}\\
p^{\text{cap}}_w(\dfrac{v_t}{v_{\text{rated}}})^3,& v_{\text{cutin}} \leq v_t \leq v_{\text{rated}}\\
0,& \text{otherwise}
\end{cases}
\end{equation}
\begin{equation}
\overline{p}_t^s = p^{\text{cap}}_s \eta^s (\frac{I_t}{I_s})
\end{equation}
where $v_t$ denotes the wind speed at stage $t$, $v_{\text{cutin}}$ denotes the cut-in speed, $v_{\text{cutout}}$ denotes the cut-out speed, $v_{\text{rated}}$ denotes the rated speed, $p^{\text{cap}}_w$ and $p^{\text{cap}}_s$ denote the wind capacity and solar capacity, $\eta^s$ denotes generation efficiency, $I_t$ and $I_s$ denote the current and standard solar radiation intensity.

For the system dynamics of energy storage, there is
%\begin{equation}
%\label{HESCapacityEq}
%\kappa_{t+1} =
%\begin{cases}
%\max \{\kappa_{t}-h_t / \eta^{\text{H2P}}, 0 \}, &\text{if} \ h_t \geq 0\\
%\min \{ \kappa_{t}-h_t \eta^{\text{P2H}}, \kappa^{\text{cap}} \}, &\text{if} \ h_t \leq 0
%\end{cases}
%\end{equation}
%where $\kappa_{t}$ is the stored hydrogen of storage in the charging station at stage $t$, $\kappa_{\text{cap}}$ is the hydrogen storage capacity of storage, $\eta^{\text{H2P}}$ is the round-trip efficiency from hydrogen to power, $\eta^{\text{P2H}}$ is the round-trip efficiency from power to hydrogen, $h_t$ is the discharge power of storage if $h_t \geq 0$, otherwise is the charge power of storage. Considering $\kappa_{t}^e = \kappa_{t} \sigma_{H_2}$ where $\kappa_{t}^e$ is the stored hydrogen energy and $\sigma_{H_2}$ is the lower heating of hydrogen, equation (\ref{HESCapacityEq}) can be rewritten as follows by multiplying both sides with $\sigma_{H_2}/\kappa_{e}^{\text{cap}}$
\begin{equation}
\label{EQHESSOCTransfer}
b_{t+1} =
\begin{cases}
\max \{b_{t}-h_t \Delta t / (\eta^{dc} \kappa_{e}^{\text{cap}}), 0 \}, &\text{if} \ h_t \geq 0\\
\min \{b_{t}-h_t \eta^{c} \Delta t / \kappa_{e}^{\text{cap}}, 1 \}, &\text{if} \ h_t \leq 0
\end{cases}
\end{equation}
where $\kappa_{e}^{\text{cap}}$ denotes the energy capacity of storage, $\eta^{dc}$ denotes the discharge efficiency of storage and $\eta^{c}$ denotes the charge efficiency of storage. $h_t$ is the discharge power of storage if $h_t \geq 0$, otherwise is the charge power of storage.

\textit{4) Constraints:} First, there is a upper and lower bound for the pricing of EV charging, i.e.,
\begin{equation}
\label{EQPriceBound}
0 \leq \varphi_t \leq \overline{\varphi}
\end{equation}
Second, the actual utilization of the distributed renewable energy should not exceed the maximum generation, i.e.,
\begin{equation}
\label{EQREBound}
0 \leq p_t^w \leq \overline{p}_t^w, \quad 0 \leq p_t^s \leq \overline{p}_t^s
\end{equation}
where $p_t^w$ and $p_t^s$ denote the actual utilization of the distributed wind energy and solar power. Third, the charging of EVs has the following constraints:
\begin{equation}
\label{EQEVBound}
0 \leq n_t + n_t^{\text{in}} \leq N, \quad n_t \geq 0, \quad n_t^{\text{in}} \geq 0
\end{equation}
\begin{equation}
\label{EQActionBound}
z_t^i \leq \tau_t^i, \quad z_t^i \in \{0,1\}
\end{equation}
\begin{equation}
\label{EQEnergyBound}
0 \leq e_t^i \leq \tau_t^i p \eta^{\text{ev}} \Delta t
\end{equation}
\begin{equation}
\label{EQChargingPower}
p_t^{\text{ev}} = \sum_{i=1}^{n_t + n_t^{\text{in}}} z_t^i p
\end{equation}
where constraint (\ref{EQEVBound}) denotes the number of parked EVs should not exceed the capacity of the charging station, constraint (\ref{EQActionBound}) regulates the charging action may happen only when there is a EV connected to the charging pile, constraint (\ref{EQEnergyBound}) guarantees each EV is charged with its requested energy, and constraint (\ref{EQChargingPower}) denotes the total charging power of the charging station. Fourth, the storage has the following constraints:
\begin{equation}
\label{EQSOC}
0 \leq b_t \leq 1
\end{equation}
\begin{equation}
\label{EQHESPower}
\overline{h}_t^c \leq h_t \leq \overline{h}_t^{\text{dc}}
\end{equation}
where $\overline{h}_t^c = -\min (h^{\text{cap}},(1-b_t) \kappa_{e}^{\text{cap}}/\eta^{\text{c}})$, $\overline{h}_t^{\text{dc}} = \min (h^{\text{cap}},b_t \kappa_{e}^{\text{cap}}\eta^{\text{dc}})$ and $h^{\text{cap}}$ denotes the maximum charge/discharge power of storage. Constraint (\ref{EQSOC}) denotes the upper and lower bound of SOC of storage and constraint (\ref{EQHESPower}) denotes the upper and lower bound of the output power of storage at stage $t$.

\textit{5) Objective Function:} The social welfare of the charging station is chosen as the one-step reward function,
\begin{equation}
\label{EQOneStepReward}
\begin{split}
\vartheta_t(s_t,a_t) = & \sum_{i=1}^{n_t + n_t^{\text{in}}} \varphi_t^i z_t^i p \Delta t - \gamma_t^g g_t -  \gamma_t^h |h_t| - \gamma_t^w p_t^w - \gamma_t^s p_t^s  \\
 & - \gamma_t^l n_t^l - \beta [\varphi_t - \dfrac{1}{T_w} \mathbf{E}\sum_{\tau = t}^{t+T_w-1} \varphi_{\tau}]^2
\end{split}
\end{equation}
where $g_t$ denotes the procure power from the power grid, $n_t^l$ denotes the number of arrival EVs which refuse to enter into the charging station, $T_w$ denotes the size of the time window, $\beta$ denotes the weighting parameter, and $\gamma_t^g$/$\gamma_t^h$/$\gamma_t^w$/$\gamma_t^s$/$\gamma_t^l$ denote the unit cost. The first term in the right equation denotes the charging service earning of the charging station. The second term denotes the procure cost from the power grid in case of insufficient supply where $g_t = \max (p_t^{\text{ev}} - p_t^w - p_t^s - h_t, 0)$. The third term denotes the operation cost of storage. The fourth and fifth terms denote the operation cost of distributed wind power and solar power. The sixth term denotes the QoS cost of the charging station where the number of EVs that refuse to enter into the charging station is used to denote the service dropping rate. The last term denotes the price fluctuation within a sliding window which is the main focus of the EV drivers.

As the prediction error increases with the time horizons\cite{foley2012current}, the expected total social welfare within a sliding time window is chosen as the objective function, i.e.,
\begin{equation}
\label{EQObj}
J(\pi,s_t) = \mathbf{E}^{\pi} [\sum_{\tau=t}^{t+T_w-1} \vartheta_\tau(s_{\tau},a_{\tau}) | s_t]
\end{equation}
where the initial state is $s_t$, the scheduling policy is $\pi$ and $\pi(s_{\tau})=a_{\tau}$. Finally, the joint scheduling of pricing and charging control can be summarized as follows which is described as problem \textbf{P1}. The objective is to find the scheduling policy that maximize the expected total social welfare defined in (\ref{EQObj}) while satisfying the constraints introduced above.
\begin{equation}
\label{EQProblem}
\begin{split}
\textbf{P1}: & \; \max_{\pi \in \Pi} J(\pi,s_t) \\
& \quad \textit{s.t.} \\
& \quad (\ref{EQPriceDiscount})-(\ref{EQHESPower})
\end{split}
\end{equation}
where $\Pi$ is the policy space.

Motivated by model predictive control \cite{wytock2017dynamic}, the energy operator in the station will solve this multi-stage stochastic programming at each stage and only implement the current decision. Observing problem \textbf{P1}, it can be found that there exists a variance computation in the reward function. Due to the glimpse into the future in the one-step reward function (\ref{EQOneStepReward}), the traditional solution approaches, such as policy iteration and value iteration, can not be directly applied \cite{xia2020risk}. In the next section, we will explore a two-phase on-line event-based optimization approach to approximately solve the problem.

\section{Solution Methodology}

\subsection{Event Definition and Performance Difference}
In order to avoid the exponentially increasing state space with the increasing scale of charging station, the event-based optimization (EBO) framework is proposed to focus on the event-triggered decision rather than the state-triggered decision which can save computation overhead \cite{Cao2007Book}. In EBO, the event is defined as the set of state transitions $<s_{t-1},s_{t}>$ with certain common properties. Therefore, the density index $n_t/N$ of the charging station is represented as the common property and the event space is defined as follows
\begin{equation}
\mathcal{E} =\{ e_t^k | t=1,2,...,T, k=1,2,3,4,5\}
\end{equation}
where
\begin{equation}
\label{EQEventDefination}
\begin{split}
e_t^1 & = \{ <s_{t-1},s_{t}> | n_t/N \in [0,0.2]\}   \\
e_t^2 & = \{ <s_{t-1},s_{t}> | n_t/N \in (0.2,0.4]\}   \\
e_t^3 & = \{ <s_{t-1},s_{t}> | n_t/N \in (0.4,0.6]\}   \\
e_t^4 & = \{ <s_{t-1},s_{t}> | n_t/N \in (0.6,0.8]\}   \\
e_t^5 & = \{ <s_{t-1},s_{t}> | n_t/N \in (0.8,1.0]\}.
\end{split}
\end{equation}
In (\ref{EQEventDefination}), $e_t^1$ and $e_t^2$ denote the density of the charging station at stage $t$ is low and weak low, respectively. $e_t^3$ denotes the density is medium, while $e_t^4$ and $e_t^5$ denote the density is weak high and high, respectively. Note that the event space can consist of any number of events. We use five events for a proof of concept here. It can be seen that the event space is fixed and far smaller than the state space in problem \textbf{P1}.

With the definition of the event, we focus on the searching of the optimal event-based scheduling policy $\sigma^{*}$ for the proposed problem, i.e., $\sigma^{*}(e_t^k)=a_{t}$. However, the traditional dynamic programming method cannot be applied due to the aforementioned difficulty of glimpsing into the future in the one-step reward function. Therefore, we will next develop a event-based policy iteration method for problem \textbf{P1}.

Based on the theory of the sensitivity-based optimization \cite{Cao2007Book}, we first derive the performance difference for two event-based scheduling policies which will be used in the event-based policy iteration. For simplicity, the following denotations are used,
\begin{equation}
\label{EQEventReward}
\begin{split}
r_{t,\sigma} = & \sum_{i=1}^{n_t + n_t^{\text{in}}} \varphi_t^i z_t^i p \Delta t - \gamma_t^g g_t -  \gamma_t^h |h_t| - \gamma_t^w p_t^w \\
 & - \gamma_t^s p_t^s  - \gamma_t^l n_t^l
\end{split}
\end{equation}
\begin{equation}
\label{EQEventCost}
c_{t,\sigma} = \varphi_t
\end{equation}
\begin{equation}
\label{EQEventAver}
\overline{J}_{\sigma} = \dfrac{1}{T_w} \mathbf{E}\sum_{\tau = t}^{t+T_w-1} c_{\tau,\sigma}
\end{equation}
where (\ref{EQEventReward}) denotes the one-step reward at stage $t$ under event-based policy $\sigma$ without considering the variance in (\ref{EQOneStepReward}), (\ref{EQEventCost}) denotes the current charging price under event-based policy $\sigma$ and (\ref{EQEventAver}) denotes the average charging price within sliding window under event-based policy $\sigma$. Based on the above denotations, (\ref{EQOneStepReward}) under event-based policy $\sigma$ can be rewritten as follows,
\begin{equation}
\vartheta_t(s_t,a_t) = r_{t,\sigma} - \beta(c_{t,\sigma}-\overline{J}_{\sigma})^2.
\end{equation}

Let $J_{t,\sigma}(e_t)$ denotes the value function from stage $t$ to $t+T_w-1$ under event-based policy $\sigma$ when observing $e_t$, $q_\sigma(e_\tau)$ denotes the probability of occurrence for $e_\tau$ under $\sigma$, $q_\sigma({s_\tau }|{e_\tau })$ denotes the probability of occurrence for $s_\tau$ under $\sigma$ when observing $e_t$ and $p_{\tau ,\sigma }$ denotes the state transition probability at stage $\tau$ under $\sigma$. Then, there is the following,
\newtheorem*{lem1}{Lemma 1}
\begin{lem1}
The performance difference of the event-based scheduling policies $\sigma$ and $\nu$ satisfies
\begin{equation}
\label{PerformanceDiff}
\begin{split}
J_{t,\sigma}(e_t) - J_{t,\upsilon}(e_t) = & \sum\limits_{\tau  = t}^{t + T_w - 1}  \sum\limits_{{e_\tau} \in \mathcal{E}} \sum\limits_{{{\rm{s}}_\tau } \in \mathcal{S}} q_\sigma(e_\tau)q_\sigma({s_\tau }|{e_\tau }) \\
& \{  [{r_{\tau ,\sigma }} - \beta {{({c_{\tau ,\sigma }} - {{\bar J}_\sigma })}^2} + {p_{\tau ,\sigma }}{J_{\tau  + 1,\upsilon }}] \\
& -  [{r_{\tau ,\upsilon }} - \beta {{({c_{\tau ,\upsilon }} - {{\bar J}_\upsilon })}^2} + {p_{\tau ,\upsilon }}{J_{\tau  + 1,\upsilon }}] \}
\end{split}
\end{equation}
\end{lem1}

The proof is given in Appendix A. Based on the performance difference equation (\ref{PerformanceDiff}), we can further obtain the sufficient condition for policy improvement, i.e.,
\newtheorem*{lem2}{Theorem 1}
\begin{lem2}
$\forall \tau = t,t+1,...,t+T_w-1$ and $\forall e_{\tau} \in \mathcal{E}$, if
\begin{equation}
\label{eventperformanceImprovement}
\begin{split}
\sum\limits_{{{\rm{s}}_\tau } \in \mathcal{S}} q_\sigma({s_\tau }|{e_\tau }) [{r_{\tau ,\sigma }} - \beta {{({c_{\tau ,\sigma }} - {{\bar J}_\upsilon })}^2} + {p_{\tau ,\sigma }}{J_{\tau  + 1,\upsilon }}] \\
 \geq \sum\limits_{{{\rm{s}}_\tau } \in \mathcal{S}} q_\sigma({s_\tau }|{e_\tau }) [{r_{\tau ,\upsilon}} - \beta {{({c_{\tau ,\upsilon}} - {{\bar J}_\upsilon })}^2} + {p_{\tau ,\upsilon }}{J_{\tau  + 1,\upsilon }}]
\end{split}
\end{equation}
, then there is $J_{t,\sigma}(e_t) \geq J_{t,\upsilon}(e_t)$. If there exists $e_{\tau^{'}}, \tau^{'} = t,t+1,...,t+T_w-1$ which the inequality strictly holds, there is $J_{t,\sigma}(e_t) > J_{t,\upsilon}(e_t)$.
\end{lem2}

The proof is given in Appendix B. Furthermore, the optimal event-based scheduling policy $\sigma^*$ satisfies the following property,
\newtheorem*{lem3}{Theorem 2}
\begin{lem3}
For any policy $\upsilon$ and the current observed event $e_t$, there is
\begin{equation}
\label{EquationOptimalPolicy}
\begin{split}
& \sum\limits_{{{\rm{s}}_t } \in \mathcal{S}} q_\sigma({s_t }|{e_t }) [{r_{t,\upsilon}} - \beta {{({c_{t ,\upsilon }} - {{\bar J}_{\sigma^*} })}^2} + {p_{t ,\upsilon }}{J_{t  + 1,\sigma^* }}] \\
& \leq \sum\limits_{{{\rm{s}}_t } \in \mathcal{S}} q_\sigma({s_t }|{e_t }) [{r_{t ,\sigma^*}} - \beta {{({c_{t ,\sigma^*}} - {{\bar J}_\sigma^* })}^2} + {p_{t ,\sigma^* }}{J_{t  + 1,\sigma^* }}]
\end{split}
\end{equation}
\end{lem3}

The proof is given in Appendix C. Based on the above performance difference and the policy improvement property of the event-based scheduling policies, we will introduce a two-phase on-line joint scheduling algorithm in the next section.

\subsection{Two-phase On-line Joint Scheduling Algorithm}

The motivation of the two-phase on-line joint scheduling algorithm is to decompose the action to avoid large action space and handle the non-linear stochastic arrivals (\ref{EQArrival}). At each time, the charging station will firstly update the pricing policy by generating the sample paths of future EV arrivals and choose the current optimal charging price (\textbf{phase I}). Then, the charging station will implement scenario-based MPC for the smart control of parked EVs and storage considering the updated pricing policy (\textbf{phase II}). The details are introduced below.

\subsubsection{Phase I On-line Policy Iteration for Pricing}

For current observed event $e_t$ and pricing policy $\sigma^k$ at stage $t$ where $k$ denotes the iteration index, the charging station will update its pricing policy and obtain the improved policy $\sigma^{k+1}$ by using the following mechanism,
\begin{equation}
\label{EQPolicyUpdate}
\begin{split}
& \sigma^{k+1} ({e_t}): = \mathop {\arg \max }\limits_{{\varphi_t}} \sum\limits_{{s_t} \in \mathcal{S}} {q_{\sigma^{k+1}} }({s_t}|{e_t})\{ {r_{t,\sigma^{k+1} }} + {p_{t,\sigma^{k+1} }}{J_{t + 1,\sigma^{k} }} \\
& \qquad \qquad \qquad \qquad \qquad \qquad - \beta {{({c_{t,\sigma^{k+1} }} - {{\bar J}_{\sigma^{k}} })}^2}\}  \\
& \sigma^{k+1} ({e_\tau }): = \sigma^{k} ({e_\tau }),\forall \tau  = t + 1,...,t + T_w - 1
\end{split}
\end{equation}
Based on Theorem 1, it can be deduced that $J_{t,\sigma^{k+1}}(e_t) \geq J_{t,\sigma^{k}}(e_t)$. In fact, equation (\ref{EQPolicyUpdate}) gives the on-line policy iteration formula for the finite-stage event-based optimization problem with variance criterion.

In order to solve (\ref{EQPolicyUpdate}), we propose a simulation-based method to estimate ${q_{\sigma^{k+1}} }({s_t}|{e_t})$, ${p_{t,\sigma^{k+1} }}{J_{t + 1,\sigma^{k} }}$ and ${{\bar J}_{\sigma^{k} }}$ by using sample paths. For current stage $t$ and observed event $e_t$, we can generate $M$ sample paths by using mixed policy $(\sigma_1^{k+1},\sigma_2^{k+1},...,\sigma_{t-1}^{k+1},\varphi_t,\sigma_{t+1}^{k},...,\sigma_{t+T_w-1}^{k})$ for current action which needs to be evaluated. Each sample path is denoted as $s_{\tau}^m, m=1,2,...,M$ and its uncertain variables $n_{\tau}^{\text{in}}$, $e_{\tau}^i$ and $\tau_{\tau}^i$ for the newly arrival EVs and $p_{\tau}^w$/$p_{\tau}^s$ for distributed wind/solar power are sampled based on their probability distributions. Therefore, there are
\begin{equation}
\label{EQQApprox}
{q_{\sigma^{k+1}}}({s_t} = s|{e_t} = e) \approx \frac{{\sum\limits_{m = 1}^M {{\bf{1}}(s_t^m = s){\bf{1}}(e_t^m = e)} }}{{\sum\limits_{m = 1}^M {{\bf{1}}(e_t^m = e)} }}
\end{equation}
\begin{equation}
\label{EQAverPriceApprox}
{\bar J_{\sigma^{k}} } \approx \frac{{\sum\limits_{m = 1}^M {{\bf{1}}(s_t^m = s){\bf{1}}(e_t^m = e)\sum\limits_{\tau  = t}^{t + T_w - 1} {{c_{\tau ,{\sigma^{k}} }}(s_\tau ^m,{\sigma^{k}} (e_\tau ^m))} } }}{{T\sum\limits_{m = 1}^M {{\bf{1}}(s_t^m = s){\bf{1}}(e_t^m = e)} }}
\end{equation}
\begin{equation}
\label{EQFutureCost}
\begin{split}
& {p_{t,{\sigma^{k+1}} }}{J_{t + 1,{\sigma^{k}} }} = {\rm{E}}\{ {J_{t + 1,{\sigma^{k}} }}({s_{t + 1}})|{e_t} = e,{s_t} = s,{\varphi_t} = \varphi\} \\
& \approx \frac{\parbox{8cm}{$\sum\limits_{m = 1}^M {\bf{1}}(s_t^m = s){\bf{1}}(e_t^m = e) \sum\limits_{\tau  = t + 1}^{t + T_w - 1} \{ {r_{\tau ,{\sigma^{k}} }}(s_\tau ^m,{\sigma^{k}} (e_\tau ^m)) - \beta {{({c_{\tau ,{\sigma^{k}} }}(s_\tau ^m,{\sigma^{k}} (e_\tau ^m)) - {{\bar J}_{\sigma^{k}} })}^2}|{\varphi_t} = \varphi\}$}}{{\sum\limits_{m = 1}^M {{\bf{1}}(s_t^m = s){\bf{1}}(e_t^m = e)} }}
\end{split}
\end{equation}
where $e_t^m$ denotes the observed event in the $m$th sample path at stage $t$, $\bf{1}(\cdot)$ denotes the indicator function and $\varphi$ denotes the pricing action which needs to be evaluated at stage $t$. Note that the sample paths before stage $t$ can be shared by all the actions to be evaluated and $q_{\sigma^{k+1}}({s_t} = s|{e_t} = e)$ only depends on the policy series $(\sigma_1^{k+1},\sigma_2^{k+1},...,\sigma_{t-1}^{k+1})$. Furthermore, the accumulated reward $\sum\limits_{\tau  = t + 1}^{t + T_w - 1}  {r_{\tau ,{\sigma^{k}} }}(s_\tau ^m,{\sigma^{k}}(e_\tau ^m))$ in (\ref{EQFutureCost}) can be determined in phase II as shown below.

\subsubsection{Phase II Scenario-based MPC for Smart Charging}

After the charging station updates the pricing policy and posts the current charging price, the station should implement smart control for the EVs and storage. This control can be formulated as a multi-stage stochastic programming problem considering the limited battery capacity and uncertain EV arrivals in the future. Therefore, we propose a scenario-based MPC for the smart charging of EVs, i.e.

\begin{equation}
\label{ScenarioMPC}
\begin{split}
& \mathop {\max } \frac{1}{M} \sum\limits_{m  = 1}^{M} \sum\limits_{\tau  = t}^{t + T_w - 1}\{ \sum\limits_{i = 1}^{{n_\tau } + {n_{\tau ,in}}} \varphi_\tau ^{i,m} z_\tau ^{i,m} p \Delta t - \gamma _\tau ^g{g_\tau^m } - \gamma_\tau ^h|{h_\tau^m }|\\
& \qquad \qquad \qquad \qquad - \gamma_\tau ^w p_\tau ^{w,m} - \gamma_\tau ^s p_\tau ^{s,m}   \}\\
& s.t. \\
& \quad   {\varphi _t^m} = \sigma_{t}^{k+1}(e_t)\\
& \quad   {\varphi _{\tau}^m} = \sigma_{\tau}^{k}(e_{\tau}^m)\\
& \quad   (\ref{EQRemainParkTime})-(\ref{EQHESSOCTransfer}), (\ref{EQREBound})-(\ref{EQHESPower}) \enspace \text{for each scenario} \enspace m\\
& \quad \forall m, (z_t^{i,m},g_t^m,h_t^m,p_t^{w,m},p_t^{s,m}) \enspace \text{is equal at stage} \enspace t
\end{split}
\end{equation}
where the superscript $m$ denotes the index of the scenario in the rest of paper. Note that it is unnecessary to generate new scenarios for problem (\ref{ScenarioMPC}). For current action ${\varphi _t^m} = \sigma_{t}^{k+1}(e_t)$, we have already generated sample paths for action evaluation in phase I. Therefore, we can directly use these sample paths as scenarios for problem (\ref{ScenarioMPC}).

It can be found that problem (\ref{ScenarioMPC}) is actually a multi-stage determined programming problem. Furthermore, this problem can be transformed into a mixed integer programming problem by linearization. By introducing auxiliary variables $z_{c,\tau}^m$/$z_{dc,\tau}^m \in \{0,1\}$ to denote the charging/discharging decisions for storage where $0 \leq z_{c,\tau}^m + z_{dc,\tau}^m \leq 1$, $h_{c,\tau}^m$/$h_{dc,\tau}^m$ to denote the charging/discharging power of storage and $y_{c,\tau}^m$/$y_{dc,\tau}^m$ to denote the temporary variables, the storage dynamic (\ref{EQHESSOCTransfer}) can be replaced by the following linear expressions,
\begin{equation}
\left\{ {\begin{array}{*{20}{l}}
{b_{\tau {\rm{ + }}1}^m{\rm{ = }}b_\tau ^m + y_{c,\tau }^m{\eta ^c}/\kappa^{{\rm{cap}}}_e - y_{dc,\tau }^m/({\eta ^{dc}}\kappa^{{\rm{cap}}}_e)}\\
h_{c,\tau }^m - {h^{{\rm{cap}}}}(1 - z_{c,\tau }^m) \le y_{c,\tau }^m \le h_{c,\tau }^m\\
h_{dc,\tau }^m - {h^{{\rm{cap}}}}(1 - z_{dc,\tau }^m) \le y_{dc,\tau }^m \le h_{dc,\tau }^m\\
0 \le y_{c,\tau }^m \le z_{c,\tau }^m{h^{{\rm{cap}}}},\quad 0 \le y_{dc,\tau }^m \le z_{dc,\tau }^m{h^{{\rm{cap}}}}
\end{array}} \right.
\end{equation}
Similarly, the constraint (\ref{EQHESPower}) on the output power of storage can be linearized by introducing the temporary variables $f_{c,\tau}^m$/$f_{dc,\tau}^m$,
\begin{equation}
\left\{ {\begin{array}{*{20}{l}}
h_{\tau}^m = h_{dc,\tau }^m - h_{c,\tau }^m\\
0 \le h_{c,\tau}^m \le z_{c,\tau}^m{h^{{\rm{cap}}}}, \quad 0 \le h_{dc,t}^m \le z_{dc,t}^m{h^{{\rm{cap}}}}\\
b_\tau ^m - (1 - z_{c,\tau }^m) \le f_{c,\tau }^m \le b_\tau ^m\\
b_\tau ^m - (1 - z_{dc,\tau }^m) \le f_{dc,\tau }^m \le b_\tau ^m\\
- \kappa^{{\rm{cap}}}_{\rm{e}}(z_{c,\tau }^m - f_{c,\tau }^m)/{\eta ^c} \le h_\tau ^m \le f_{dc,\tau }^m\kappa^{{\rm{cap}}}_{\rm{e}}{\eta ^{dc}}\\
0 \le f_{c,\tau }^m \le z_{c,\tau }^m, \quad 0 \le f_{dc,\tau }^m \le z_{dc,\tau }^m
\end{array}} \right.
\end{equation}
For $|{h_\tau^m }|$ and $g_\tau^m$ in objective function (\ref{ScenarioMPC}), it can be linearized by introducing auxiliary variables $u_{\tau}^m$/$v_{\tau}^m$ and $z_{u,\tau}^m$/$z_{v,\tau}^m$. There are
\begin{equation}
|{h_\tau^m }| = u_{\tau}^m - v_{\tau}^m, u_{\tau}^m>0, v_{\tau}^m >0
\end{equation}
\begin{equation}
\left\{ {\begin{array}{*{20}{l}}
0 \leq g_\tau^m, \quad p_{\tau}^{ev,m} - p_{\tau}^{w,m} - p_{\tau}^{s,m} - h_{\tau}^m \leq g_\tau^m\\
g_\tau^m - M(1-z_{u,\tau}^m) \leq 0\\
g_\tau^m - M(1-z_{v,\tau}^m) \leq p_{\tau}^{ev,m} - p_{\tau}^{w,m} - p_{\tau}^{s,m} - h_{\tau}^m\\
z_{u,\tau}^m + z_{v,\tau}^m \geq 1, z_{u,\tau}^m / z_{v,\tau}^m  \in \{0,1\}
\end{array}} \right.
\end{equation}

With these linearization mechanisms, the optimization problem in phase II can be transformed into a scenario-based mixed integer linear programming which can be solved by traditional solvers, such as Cplex, Gurobi, etc.

\subsection{Algorithm Summary}

The overall framework of the proposed two-phase on-line joint scheduling algorithm for the charging station is summarized in Algorithm 1.
\begin{algorithm}[h]
\caption{Two-phase on-line joint scheduling for charging station}
\begin{algorithmic}[1]
\STATE Initialize policy $\sigma^{k}$ where $k=0$ and set the optimal objective function $J_{t}^{*}(e_t)=0, \forall t=1,2...,T$;
\FOR{$t=1,2,...,T$}
\STATE Observe the current event $e_t$;
\FOR{each price $\varphi_t$}
\STATE Generate $M$ sample paths of distributed wind/solar power generation and EV arrivals based on $(\sigma_1^{k+1},\sigma_2^{k+1},...,\sigma_{t-1}^{k+1},\varphi_t,\sigma_{t+1}^{k},...,\sigma_{t+T_w-1}^{k})$;
\STATE Implement scenario-based MPC to compute accumulated reward $\sum\limits_{\tau  = t + 1}^{t + T_w - 1}  {r_{\tau ,{\sigma^{k}} }}(s_\tau ^m,{\sigma^{k}}(e_\tau ^m))$ based on (\ref{ScenarioMPC});
\STATE Estimate $p_{t,{\sigma^{k+1}} }J_{t + 1,{\sigma^{k}}}$ based on (\ref{EQFutureCost});
\ENDFOR
\STATE Estimate $q_{\sigma^{k+1}}({s_t}|{e_t})$ and $\bar J_{\sigma^{k}}$ based on (\ref{EQQApprox}) and (\ref{EQAverPriceApprox});
\STATE Choose the price $\varphi_t$ which has the maximal value of (\ref{EQPolicyUpdate}) and record its corresponding objective function as $J_{t}^{k}(e_t)$;
\STATE If $J_{t}^{k}(e_t) > J_{t}^{*}(e_t)$, $J_{t}^{*}(e_t) = J_{t}^{k}(e_t)$ and $\sigma_{t}^{k+1}(e_t) = \varphi_t$, else $\sigma_{t}^{k+1}(e_t) = \sigma_{t}^{k}(e_t)$;
\STATE If still in policy learning, select pricing action based on $\varepsilon$-greedy searching mechanism;
\STATE Post the selected price and observe the realized state $s_t$;
\STATE Implement scenario-based MPC for smart charging at stage $t$ based on (\ref{ScenarioMPC});
\ENDFOR
\STATE If $\sigma^{k+1} = \sigma^{k}$ or the maximum iteration $k$ is reached, the learning can be stopped, else $k=k+1$ and go to step 2.
\end{algorithmic}
\end{algorithm}

\begin{figure}
\centering
\setlength{\abovecaptionskip}{-6pt}
\includegraphics[height=0.2\textheight]{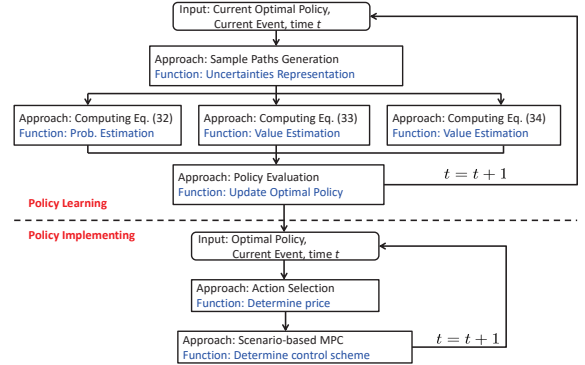}
\caption{Flowchart of the two-phase on-line joint scheduling algorithm.}
\label{FlowChart}
\end{figure}

Note that step 6 is used to compute the accumulated reward which is required for the estimation of $p_{t,{\sigma^{k+1}} }J_{t + 1,{\sigma^{k}}}$. When implementing scenario-MPC in step 6, the sample paths which has occurrence of $e_t$ will be chosen as scenarios. Step 12 and step 16 are used during policy learning. The $\varepsilon$-greedy searching mechanism can help to improve the explore capability and avoid falling into local optimum. After learning, step 12 and step 16 can be removed in practice. In fact, the proposed algorithm can be considered as asynchronous policy iteration and the policy iteration happens only for observed events. Fig. \ref{FlowChart} shows the flowchart of the proposed algorithm. In policy learning, it will iteratively find the optimal policy by simulation-based policy evaluation. In policy implementing, it will determine the price and the control scheme by using optimal policy and scenario-based MPC, respectively.

\section{Numerical Results}

\subsection{Parameter Settings}
\begin{figure}
\centering
\setlength{\abovecaptionskip}{-6pt}
\includegraphics[height=0.2\textheight]{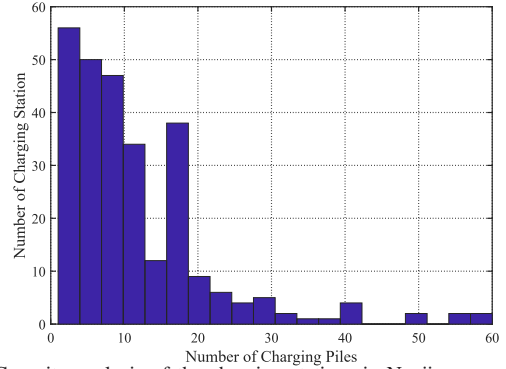}
\caption{Capacity analysis of the charging stations in Nanjing.}
\label{ChargingPileAnalysis}
\end{figure}

We have investigated the capacity of 215 charging stations in Nanjing and the result is shown in Fig. \ref{ChargingPileAnalysis}. It can be found that most of charging stations have less than 20 charging piles. Therefore, we consider a charging station with 20 charging piles in the experiments.

\begin{figure}
\centering
\setlength{\abovecaptionskip}{-6pt}
\includegraphics[height=0.2\textheight]{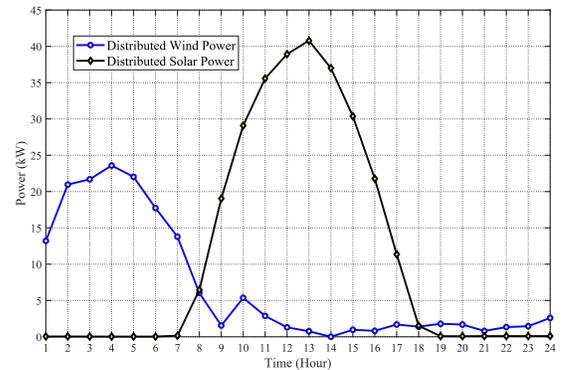}
\caption{The predicted output of distributed wind/solar power.}
\label{WindSolarPower}
\end{figure}

In the charging station, the wind speed data and solar radiation intensity come from \cite{WebsiteRenewable}. Based on the parameter settings of the distributed wind power and solar power in Table \ref{TBPara}, the predicted wind power and solar power are shown in Fig. \ref{WindSolarPower}. We assume the uncertainty of the wind/solar power follows normal distribution with the predicted value as the mean value and 10\% of the mean value as the standard deviation. The parameters of storage in the charging station are also shown in Table \ref{TBPara}.

\begin{table}[!t]
\scriptsize
\centering
\caption{\label{TBPara}Parameter Settings}
\begin{tabular}{c c c c}
\toprule
Parameter & Setting & Parameter & Setting\\
\midrule
$p^{\text{cap}}_w$ & 50kW & $p^{\text{cap}}_s$ & 50kW\\
$v^{\text{cutin}}$ & 3.5m/s & $v^{\text{cutout}}$ & 25m/s\\
$v^{\text{rated}}$ & 15m/s & $\eta^s$ & 0.88\\
$I_s$ & 800$\text{W/m}^2$ & $\beta$ & 2\\
$\kappa_{e}^{\text{cap}}$ & 166.65kWh & $h^{\text{cap}}$ & 50kW\\
$\eta^{c}$ & 0.82 & $\lambda_t$ & 10\\
$p$ & 3.6kW & $\eta^{\text{ev}}$ & 0.92\\
$\overline{\varphi}$ & 2.5CNY & $\theta_p$ & 1/25\\
$\gamma_t^w$/$\gamma_t^s$ & 0.018CNY/kWh & $\gamma_t^l$ & 1.8396CNY\\
$M$ & 100 & $\gamma_t^h$ & 0.04CNY/kWh\\
\bottomrule
\end{tabular}
\end{table}

We assume that the parking duration of EVs which enter into the charging station follows random distribution within 6 hours. The required charging energy of these EVs are sampled based on the probability distribution of the trip distance and the electric drive efficiency as introduced in \cite{HQL2017}. Other parameters considering the charging of EVs are shown in Table \ref{TBPara}.
\begin{table}[!t]
\scriptsize
\centering
\caption{\label{TBEP}Electricity Price $\gamma_t^g$}
\begin{tabular}{c c}
\toprule
Price & Time\\
\midrule
0.3208RMB/kWh & 23:00:00-6:00\\
0.8145RMB/kWh & 7:00-9:00, 15:00-16:00, 21:00-22:00\\
1.3332RMB/kWh & 13:00-14:00, 17:00-18:00\\
1.4615RMB/kWh & 10:00-12:00, 19:00-20:00 \\
\bottomrule
\end{tabular}
\end{table}

We consider the joint scheduling problem on a daily basis with $T=24$, $\Delta t = 1$hour and time window $T_w=6$hours. The unit prices of distributed wind power, solar power and storage come from \cite{long2021joint} and the detailed electricity price $\gamma_t^g$ is shown in Table \ref{TBEP}. The initial policy $\sigma^{k}$ where $k=0$ implements constant pricing scheme, i.e., $\varphi_t=2.3$CNY/kWh.

\subsection{Performance Comparison}

Firstly, we compare our method with different existing scheduling methods in practice. For the pricing scheme, we consider a high pricing scheme ($\varphi_t=2.3$CNY/kWh), a low pricing scheme ($\varphi_t=0.3$CNY/kWh) and event-driven pricing strategy\cite{xiang2021routing}, respectively. For the charging control of EVs, we consider the greedy charging scheme and delayed charging scheme, respectively. The greedy charging scheme means that EVs will be charged as soon as possible, while the delayed charging means the charging process of EVs will be postponed as long as possible. In these two charging schemes, the discharging/charging of storage and the electricity procuring will be used in turn. In this way, the following scheduling policies will be studied,

$\pi^*$: OptPrice+OptCharge. This is the proposed method in this paper.

$\pi_1$: OptPrice+GreedyCharge. This means the on-line policy iteration for pricing is used as introduced in section IV.B and the greedy charging scheme is used.

$\pi_2$: OptPrice+DelayCharge. This means the on-line policy iteration for pricing is used as introduced in section IV.B and the delayed charging scheme is used.

$\pi_3$: HighPrice+OptCharge. This means the high pricing scheme is used and the scenario-based MPC for smart charging is used.

$\pi_4$: HighPrice+GreedyCharge. This means the high pricing scheme is used and the greedy charging scheme is used.

$\pi_5$: HighPrice+DelayCharge. This means the high pricing scheme is used and the delayed charging scheme is used.

$\pi_6$: LowPrice+OptCharge. This means the low pricing scheme is used and the scenario-based MPC for smart charging is used as introduced in section IV.B.

$\pi_7$: LowPrice+GreedyCharge. This means the low pricing scheme is used and the greedy charging scheme is used.

$\pi_8$: LowPrice+DelayCharge. This means the low pricing scheme is used and the delayed charging scheme is used.

$\pi_9$: EventPrice+OptCharge. The pricing strategy comes from \cite{xiang2021routing} and the scenario-based MPC is applied to derive the optimal control scheme. This can be set as the benchmark.

The detailed performance of these policies are shown in Table \ref{TBCOST}. The second column denotes the value of the objective function. The third and fourth columns denote the operation profit and earning of the charging station, respectively. The fifth column denotes the cost of power procurement from the grid. The sixth, seventh and eighth columns denote the operation cost of the storage, distributed wind power and solar power. The ninth column denotes the QoS cost of the charging station. The eleventh, twelfth and tenth columns represent the number of arrival EVs, the number of EVs which enter into the station and their ratio. The thirteenth column denotes the price fluctuation and the last column denotes the average charging cost per EV.

\begin{table*}[!t]
\scriptsize
\centering
\caption{\label{TBCOST}Performance Comparison of the Policies (price unit: CNY)}
\begin{tabular}{c c c c c c c c c c c c c c}   %{p{0.8cm}  p{0.7cm}  p{0.7cm}   p{1.0cm}  p{0.7cm}   p{0.7cm}  p{1.3cm}  p{1.3cm}  p{1.3cm}  p{1.3cm} p{1.3cm}  p{1.3cm} p{1.3cm}}
\toprule
Policy & Obj. & Profit & Earning & Procure & \makebox[0.04\textwidth][c]{Storage Cost} & \makebox[0.05\textwidth][c]{Wind Cost} & \makebox[0.05\textwidth][c]{Solar Cost} & \makebox[0.05\textwidth][c]{QoS Cost} & \makebox[0.06\textwidth][c]{Service Ratio} & \makebox[0.06\textwidth][c]{Arrival Num.} & \makebox[0.05\textwidth][c]{Enter Num.} & \makebox[0.04\textwidth][c]{Price Std.} & \makebox[0.04\textwidth][c]{Avg. Cost}\\
\midrule
$\pi^*$ & 395.63 & 660.90 & 772.72 & 102.10 & 2.11 & 2.94 & 4.67 & 265.27 & 0.41 & 243.4 & 99.2  & 0.24 & 7.79 \\
$\pi_1$ & 351.86 & 616.03 & 755.89 & 127.92 & 5.79 & 2.77 & 3.38 & 264.16 & 0.41 & 241.9 & 98.3  & 0.22 & 7.69 \\
$\pi_2$ & 329.95 & 599.08 & 731.79 & 120.63 & 5.92 & 2.68 & 3.48 & 269.13 & 0.40 & 244.1 & 97.8  & 0.22 & 7.48 \\
$\pi_3$ & -171.94 & 238.47 & 242.96 & 0.37 & 1.84 & 1.04 & 1.23 & 410.41 & 0.08 & 242.1 & 19    & 0 & 12.79 \\
$\pi_4$ & -170.27 & 232.78 & 237.37 & 0     & 2.95 & 0.98 & 0.65 & 403.05 & 0.08 & 237.8 & 18.7  & 0 & 12.69 \\
$\pi_5$ & -170.01 & 239.12 & 243.94 & 0     & 3.16 & 1.01 & 0.64 & 409.12 & 0.08 & 241.6 & 19.2  & 0 & 12.70 \\
$\pi_6$ & -197.03 & 8.08 & 204.55 & 187.18 & 1.42 & 2.98 & 4.89 & 205.11 & 0.53 & 238.6 & 127.1 & 0 & 1.61 \\
$\pi_7$ & -259.48 & -60.44 & 211.93 & 262.50 & 2.81 & 2.95 & 4.11 & 199.04 & 0.55 & 237.9 & 129.7 & 0 & 1.63 \\
$\pi_8$ & -247.70 & -43.14 & 202.58 & 235.14 & 3.70 & 2.92 & 3.96 & 204.56 & 0.54 & 239   & 127.8 & 0 & 1.58 \\
$\pi_9$ & 364.34 & 611.03 & 727.72 & 105.32 & 3.60 & 2.98 & 4.78 & 246.68 & 0.44 & 239.9   & 105.8 & 0.34 & 6.88 \\
\bottomrule
\end{tabular}
\end{table*}

From Table \ref{TBCOST}, it can be seen that comparing with all the policies, the proposed method $\pi^*$ achieves the highest performance in the aspects of objective function, operation profit and earning of the charging station. The policies $\pi^*$, $\pi_1$ and $\pi_2$ are better than policies $\pi_3$, $\pi_4$ and $\pi_5$, while the latter policies are better than policies $\pi_6$, $\pi_7$ and $\pi_8$. This demonstrates the effectiveness of the price optimization and the high pricing scheme is better than the low pricing scheme in this experiment settings. As policies from $\pi_3$ to $\pi_8$ use constant pricing scheme, there is no pricing fluctuation during operation. Comparing $\pi^*$ with policies $\pi_3$ and $\pi_6$, it shows that the operation profit can be largely increased which demonstrates the importance of the price optimization even when the smart charging is implemented. Furthermore, it can be found that the values of the objective function from policies $\pi_3$ to $\pi_8$ are negative. For policies $\pi_3$, $\pi_4$ and $\pi_5$, this is because the high charging price will incur the potential high QoS cost due to the unacceptance of the price and the departure of the arrival EVs. For policies $\pi_6$, $\pi_7$ and $\pi_8$, although their QoS costs are low, their operation profits are also very low due to the low pricing scheme. By comparing $\pi_1$ with policies $\pi_4$ and $\pi_7$, it can be found that $\pi_1$ achieves the highest performance. Similar relationship happens for policy $\pi_2$ and policies $\pi_5$ and $\pi_8$. This further demonstrates the effectiveness of the proposed on-line policy iteration for pricing even the control of EV charging and storage is heuristic.

By comparing policy $\pi^*$ with $\pi_1$ and $\pi_2$, it can be seen that $\pi^*$ achieves the best performance. Similar relationship happens among ($\pi_3$, $\pi_4$, $\pi_5$) and ($\pi_6$, $\pi_7$ $\pi_8$), respectively. This demonstrates the effectiveness of the proposed scenario-based MPC for smart charging. The joint scheduling of pricing and charging control will maximize the operation efficiency of the charging station. From Table \ref{TBCOST}, it can be also found that the procure cost from the grid is highest for policies $\pi_6$, $\pi_7$ and $\pi_8$, while is lowest for policies $\pi_3$, $\pi_4$ and $\pi_5$. This is also caused by the pricing effect. The high price in policies $\pi_3$, $\pi_4$ and $\pi_5$ attracts less EVs to enter into the charging station, i.e., low service ratio. Thus, the charging station can be self-sufficient by using distributed renewable energy and storage. On the contrary, the low price in policies $\pi_6$, $\pi_7$ and $\pi_8$ attracts a large number of EVs to enter into the station, i.e., high service ratio. Thus, the charging demand is largely increased and the power procure cost is the highest. Based on the performance comparison of these policies, it can be seen that the optimal charging cost per EV is about 7.79CNY to attract EVs to enter into the station for charging.

By comparing policy $\pi^*$ with $\pi_9$, it can be found that the performance of $\pi^*$ is better than the benchmark policy $\pi_9$. In $\pi_9$, the pricing strategy is derived in real-time and driven by event. In addition to these, policy $\pi^*$ continuously learns the optimal pricing strategy considering the impact of the uncertainties and the control scheme in the future. This brings in the improved profit and reduced operation cost comparing policy $\pi^*$ with $\pi_9$. Furthermore, it can be seen that the price fluctuation of the policy $\pi_9$ is larger than policy $\pi^*$ due to no limitations on the price fluctuation in $\pi_9$.

\subsection{Performance Analysis}

\begin{figure}
\centering
\setlength{\abovecaptionskip}{-6pt}
\includegraphics[height=0.2\textheight]{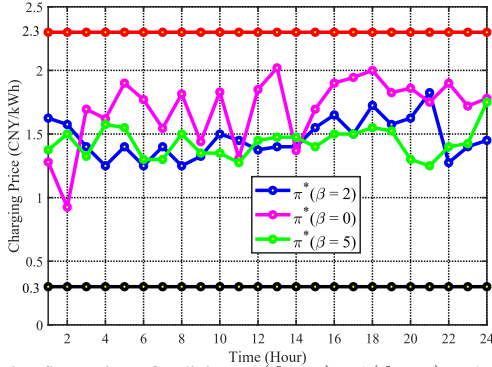}
\caption{Price fluctuation of policies $\pi^*(\beta=2)$, $\pi^*(\beta=5)$ and $\pi^*(\beta=0)$.}
\label{PriceFluctuation}
\end{figure}

Fig. \ref{PriceFluctuation} shows the pricing result by implementing policies $\pi^*$ with $\beta=2$, $\beta=0$ and $\beta=5$, respectively. The price gap for policy $\pi^*(\beta=0)$ is 1.10CNY, i.e., the highest price minus the lowest price, while the price gap for policies $\pi^*(\beta=2)$ and $\pi^*(\beta=5)$ are 0.57CNY and 0.5CNY, respectively. Furthermore, the price fluctuation for $\pi^*(\beta=0)$ is 0.51CNY, while the price fluctuation for policies $\pi^*(\beta=2)$ and $\pi^*(\beta=5)$ are 0.24CNY and 0.18CNY, respectively. Based on these results, it can be found that there exists large fluctuations for policy $\pi^*(\beta=0)$. The large fluctuation in pricing will reduce the charging willingness of the frequenters due to the uncertainties and variation of the charging cost for the EV drivers. Comparing policies $\pi^*(\beta=2)$ and $\pi^*(\beta=5)$ with $\pi^*(\beta=0)$, it demonstrates the necessity of the consideration of the last term in (\ref{EQOneStepReward}). Note that the price fluctuation will also influence the operation of the charging station. Therefore, it is suggested that the charging station obtain a suitable $\beta$ by balancing the driver's preference and the operation efficiency.

\begin{figure}
\centering
\setlength{\abovecaptionskip}{-6pt}
\includegraphics[height=0.2\textheight]{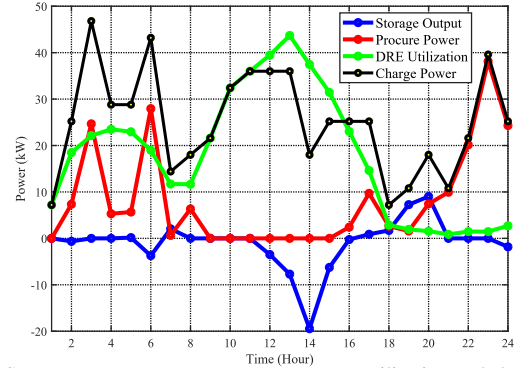}
\caption{Storage output, power procurement, DRE utilization and charge power in policy $\pi^*$.}
\label{ScenarioAnalysis}
\end{figure}

Fig. \ref{ScenarioAnalysis} shows the detailed control for storage, power procurement, distritbuted renewable energy utilization (DRE) and EV charging in policy $\pi^*$. Due to the relatively cheap price for electricity procurement, the total charging power at night is usually larger than the total charging power in the daytime. It can be found that the DRE utilization follows the trend of the total charging power most of time. The storage will charge during period (12:00-15:00) for the excess generation of the DRE, while begin to discharge during period (19:00-20:00) when the electricity price is the highest according to Table \ref{TBEP}. The electricity procurement from the grid happens when the EV charging station cannot be self-sufficient, such as during periods (2:00-6:00) and (20:00-24:00). Due to this orderly energy management, the operation efficiency of the charging station can be improved.

%Fig. \ref{PolicyStructure} shows the policy structure for pricing in policy $\pi^*$. Based on the event definition in (\ref{EQEventDefination}), the large index of the event means the high occupance of the charging station. We normalize the optimized price at each stage by the upper bound of the price. The deeper color in the figure means the higher charging price. It can be seen that the optimized price tends to be higher when the event index is large, i.e., there is few unoccupied charging piles in the station. After the on-line policy iteration, the optimized policy $\pi^*$ can be stored as a table and the price at each stage can be quickly obtained based on this look-up table.
%\begin{figure}
%\centering
%\setlength{\abovecaptionskip}{-6pt}
%\includegraphics[height=0.21\textheight]{PolicyStructure.eps}
%\caption{Policy structure for pricing in policy $\pi^*$.}
%\label{PolicyStructure}
%\end{figure}

\begin{figure}
\centering
\setlength{\abovecaptionskip}{-6pt}
\includegraphics[height=0.2\textheight]{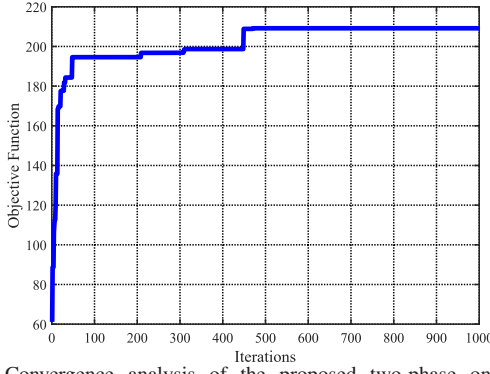}
\caption{Convergence analysis of the proposed two-phase on-line joint scheduling algorithm.}
\label{Convergence}
\end{figure}

In order to analyze the convergence of the proposed method, the value of the objective function during on-line policy iteration for event $e_4$ at 8:00 is shown in Fig. \ref{Convergence}. It can be seen that the objective function quickly improves at the beginning of the iteration and converges after about 450 iterations. Meanwhile, during policy learning, the computation time for each decision stage is about 15 seconds. This means the running time is acceptable for the joint scheduling of the charging station. These demonstrate the computation efficiency of the proposed method.

\subsection{Sensitivity Analysis}

\begin{figure*}
  \centering
  \subfigure[Social Welfare w.r.t increasing capacity]{
    \label{fig:subfig:a} %% label for first subfigure
    \includegraphics[width=0.28\textwidth]{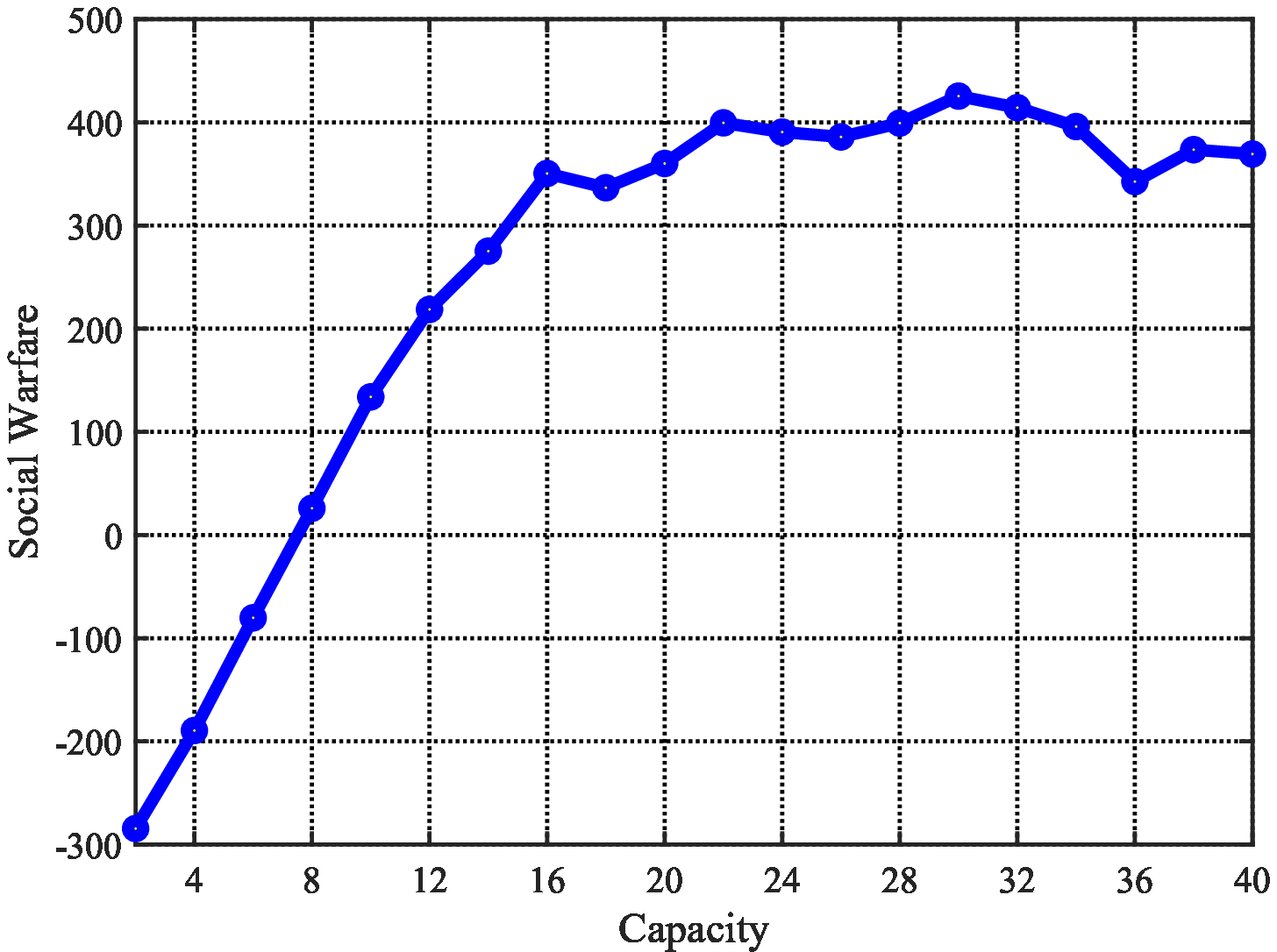}}
%  \hspace{1in}
  \subfigure[Profit w.r.t increasing capacity]{
    \label{fig:subfig:b} %% label for first subfigure
    \includegraphics[width=0.28\textwidth]{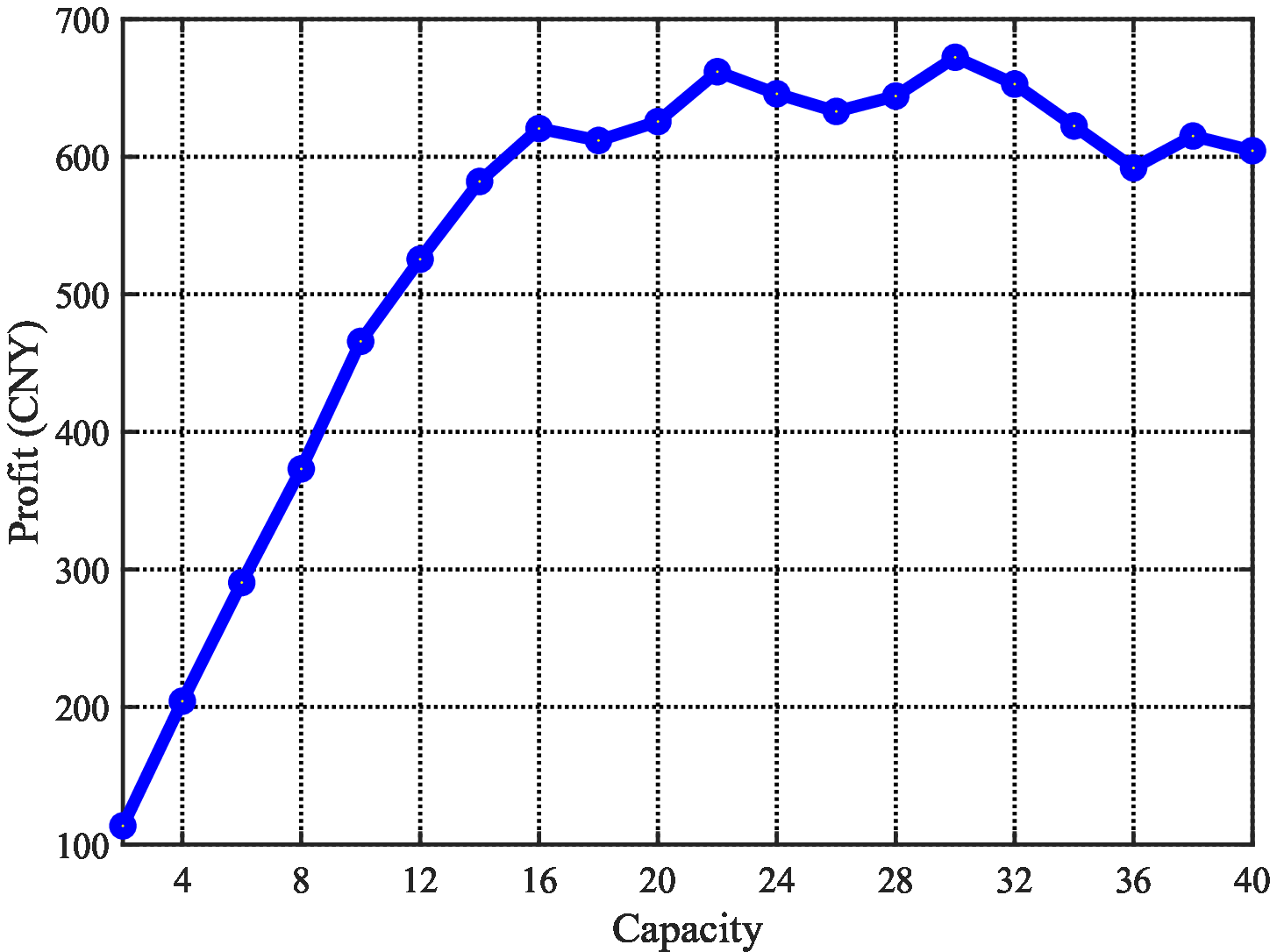}}
    %  \hspace{1in}
  \subfigure[Service Num. w.r.t increasing capacity]{
    \label{fig:subfig:c} %% label for first subfigure
    \includegraphics[width=0.28\textwidth]{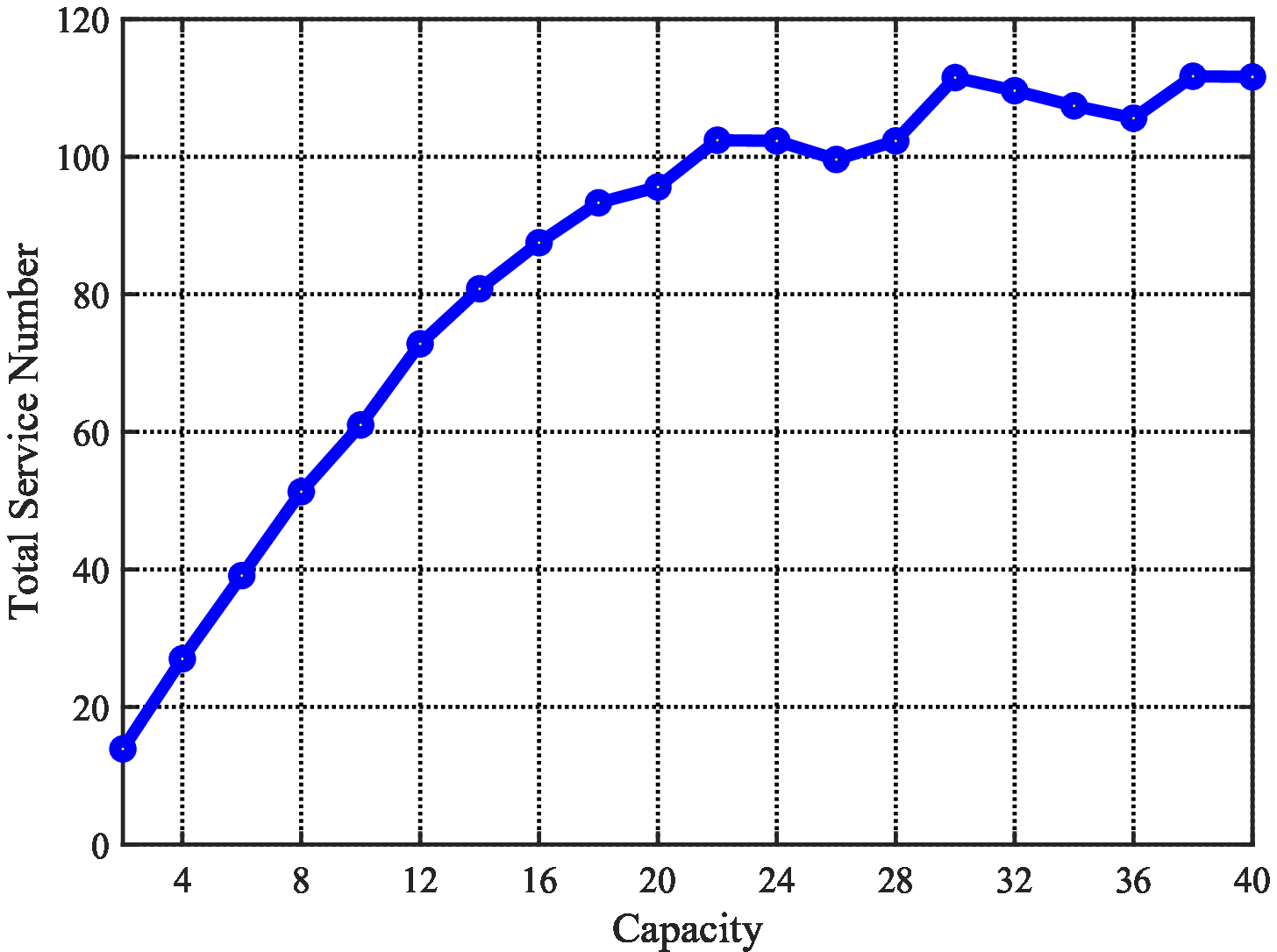}}
    %  \hspace{1in}

    \subfigure[Social Welfare w.r.t increasing arrival rate]{
    \label{fig:subfig:d} %% label for first subfigure
    \includegraphics[width=0.28\textwidth]{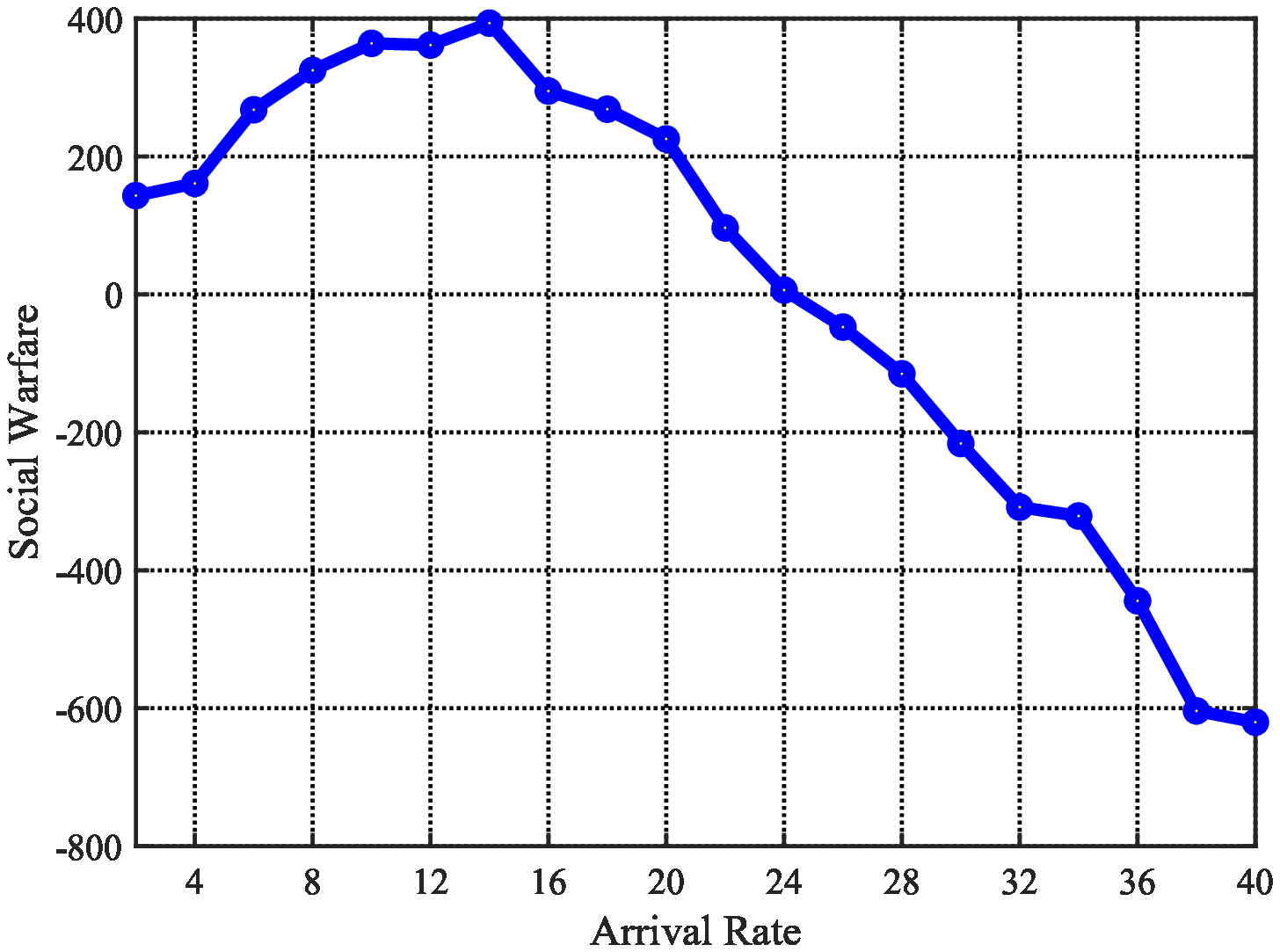}}
 % \hspace{1in}
  \subfigure[Profit w.r.t increasing arrival rate]{
    \label{fig:subfig:e} %% label for first subfigure
    \includegraphics[width=0.28\textwidth]{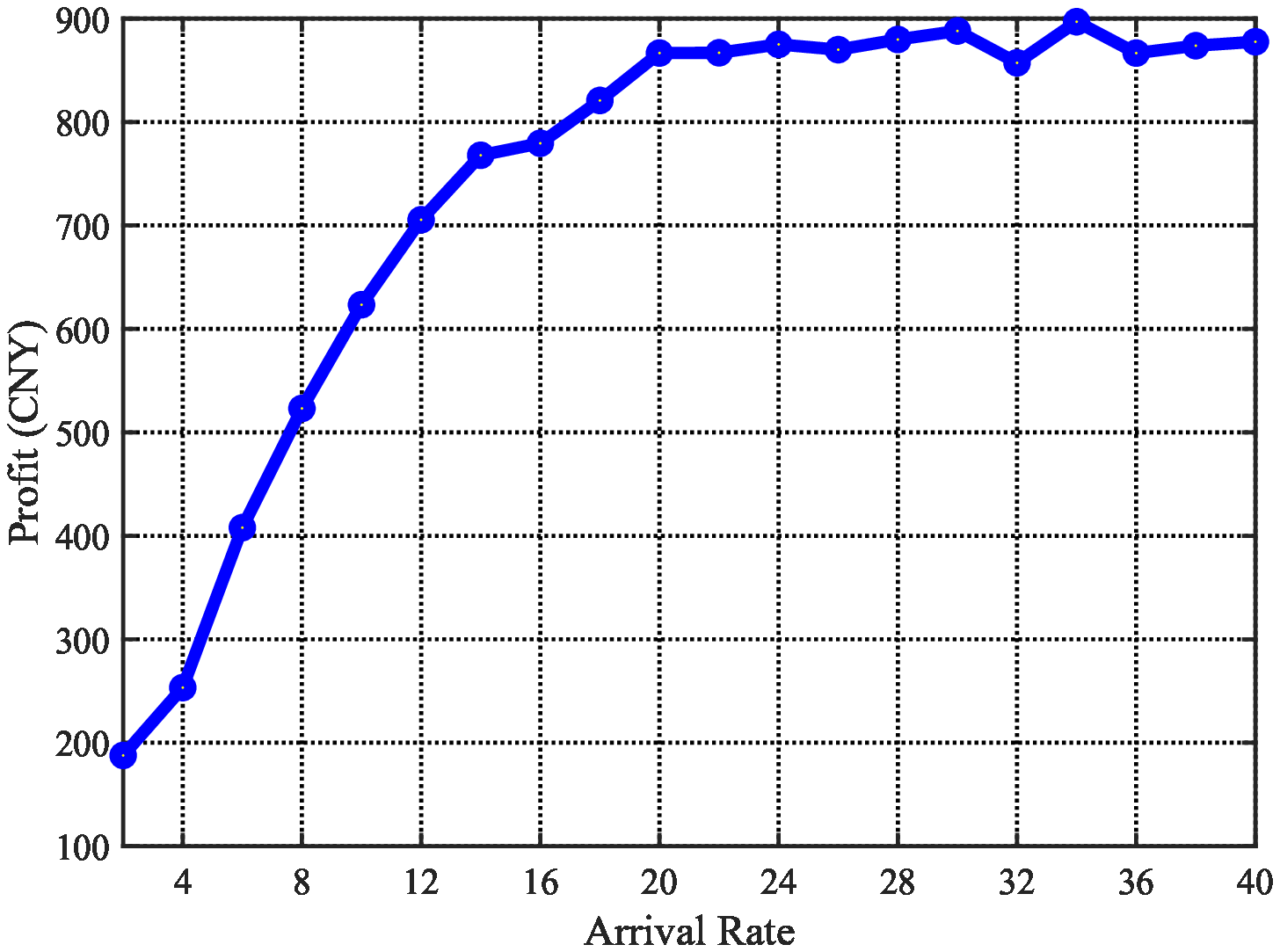}}
    %  \hspace{1in}
  \subfigure[Service Num. w.r.t increasing arrival rate]{
    \label{fig:subfig:f} %% label for first subfigure
    \includegraphics[width=0.28\textwidth]{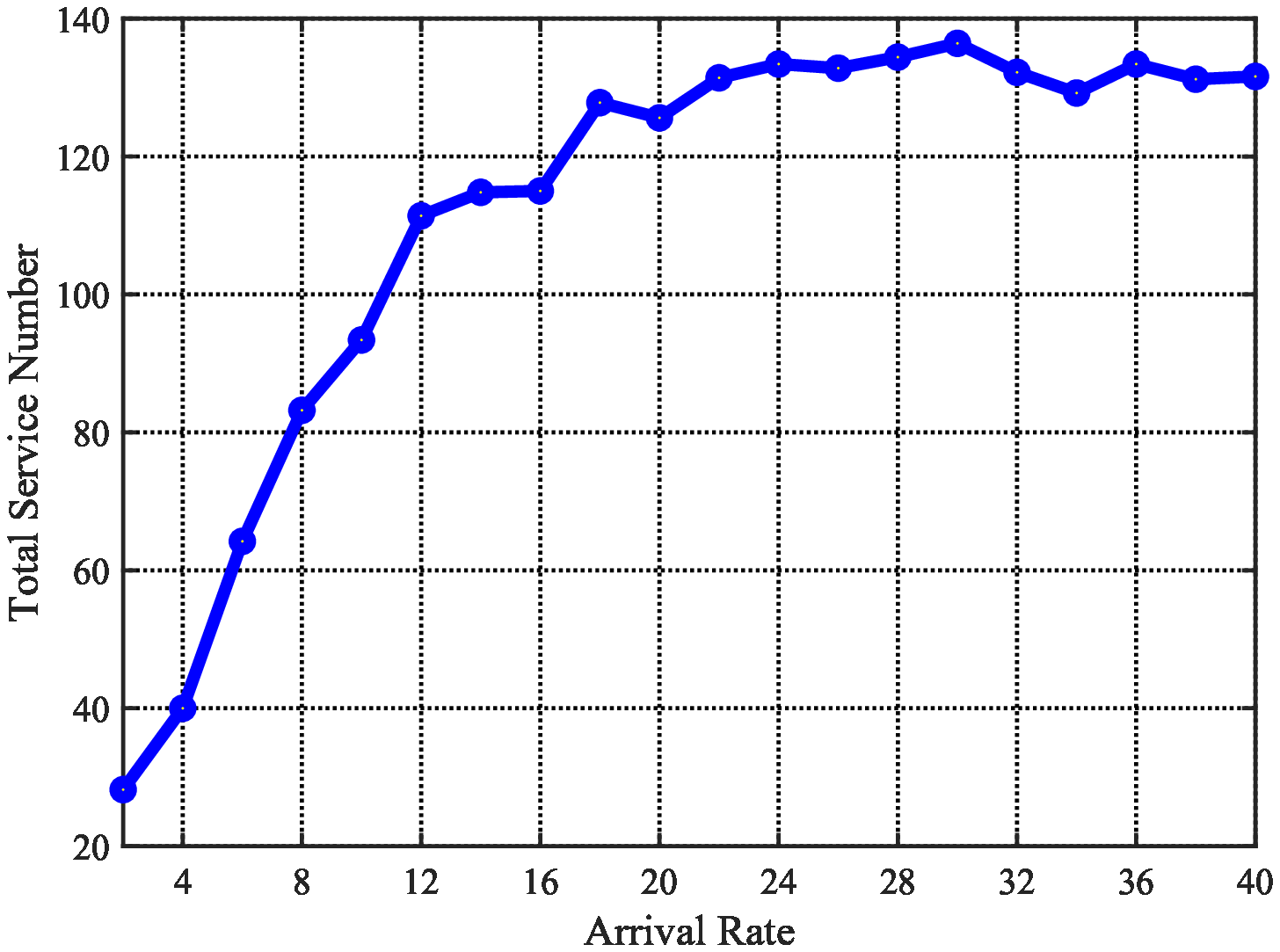}}
    %  \hspace{1in}
  \caption{Sensitivity analysis of different capacity of the charging station and the arrival rate.}
  \label{fig:subfig} %% label for entire figure
\end{figure*}

As the capacity of the charging piles and the arrival rate of EVs are critical to the operation of the charging station, we have also conducted sensitivity analysis to these factors by implementing policy $\pi^*$. The result is shown in Fig. \ref{fig:subfig}. From Fig. (a) to (c), it can be found that the social welfare, profit and the total service number of the charging station all increase with the capacity of the charging station increasing from 1 to 16. However, these curves appear to be steady when the capacity excesses 20. This is because the capacity of the charging station is enough to handle the current arrival rate $\lambda_t=10$ of EVs in this case.

From Fig. (d) to (e), it can be seen that the social welfare firstly increases and then decreases with the increasing of the arrival rate. The optimal arrival rate for the charging station with capacity $N=20$ is about 14. This is because the increasing of the arrival rate will bring in the increasing service number of EVs as depicted in Fig. (f) which will improve the operation profit of the charging station as depicted in Fig. (d). However, when the arrival rate is too large, it will excess the service capacity of the charging station and incur the large QoS cost which makes the social welfare decreased. Similarly, due to the limited service capacity, the profit and the total service number also appear to be steady after the arrival rate excesses 20. These indicate it is necessary to investigate the EV traffic before planning the capacity of the charging station.

\section{Conclusion}

In this paper, the joint scheduling problem of pricing and charging control is studied to maximize the social welfare of the charging station considering the QoS and price fluctuation sensitivity of EVs. A two-phase online policy learning method is proposed to solve the formulated MDP model in the framework of event-based optimization. By utilizing performance difference theory, the performance improvement in each policy iteration is theoretically proved. Numerical results demonstrate the effectiveness of the proposed method and the operation improvement of the charging station.

% or


\appendices

\section{Proof of Lemma 1}
\begin{proof}
\begin{equation}
\begin{split}
& J_{t,\sigma}(e_t) - J_{t,\upsilon}(e_t) \\
= &  \sum\limits_{{e_t} \in \mathcal{E}} \sum\limits_{{{\rm{s}}_t } \in \mathcal{S}} q_\sigma(e_t)q_\sigma({s_t}|{e_t}) \{  [{r_{t ,\sigma }} - \beta {{({c_{t ,\sigma }} - {{\bar J}_\sigma })}^2}+ {p_{t ,\sigma }}{J_{t  + 1,\sigma }} ]  \\
& -  [{r_{t ,\upsilon }} - \beta {{({c_{t ,\upsilon }} - {{\bar J}_\upsilon })}^2} + {p_{t ,\upsilon }}{J_{t  + 1,\upsilon }}] \}  \\
= & \sum\limits_{{e_t} \in \mathcal{E}} \sum\limits_{{{\rm{s}}_t } \in \mathcal{S}} q_\sigma(e_t)q_\sigma({s_t}|{e_t}) \{  [{r_{t ,\sigma }} - \beta {{({c_{t ,\sigma }} - {{\bar J}_\sigma })}^2}+ {p_{t ,\sigma }}{J_{t  + 1,\upsilon}} ]  \\
& -  [{r_{t ,\upsilon }} - \beta {{({c_{t ,\upsilon }} - {{\bar J}_\upsilon })}^2} + {p_{t ,\upsilon }}{J_{t  + 1,\upsilon }}]  + p_{t ,\sigma } (J_{t  + 1,\sigma } - J_{t  + 1,\upsilon }) \}\\
= & \sum\limits_{{e_t} \in \mathcal{E}} \sum\limits_{{{\rm{s}}_t } \in \mathcal{S}} q_\sigma(e_t)q_\sigma({s_t}|{e_t}) \{  [{r_{t ,\sigma }} - \beta {{({c_{t ,\sigma }} - {{\bar J}_\sigma })}^2}+ {p_{t ,\sigma }}{J_{t  + 1,\upsilon}} ]  \\
& -  [{r_{t ,\upsilon }} - \beta {{({c_{t ,\upsilon }} - {{\bar J}_\upsilon })}^2} + {p_{t ,\upsilon }}{J_{t  + 1,\upsilon }}] \} \\
& +\sum\limits_{{e_{t+1}} \in \mathcal{E}} \sum\limits_{{{\rm{s}}_{t+1} } \in \mathcal{S}}  q_\sigma(e_{t+1})q_\sigma({s_{t+1}}|{e_{t+1}}) (J_{t  + 1,\sigma } - J_{t  + 1,\upsilon }) \\
\end{split}
\end{equation}

The first equation holds as the initial distribution $q_\sigma(e_t)q_\sigma({s_t}|{e_t})$ is independent with policy $\sigma$ or $\upsilon$ and the recursion is conducted. The third equation holds as there is $\sum\limits_{{e_t} \in \mathcal{E}} \sum\limits_{{{\rm{s}}_t } \in \mathcal{S}} q_\sigma(e_t)q_\sigma({s_t}|{e_t})p_{t ,\sigma }=\sum\limits_{{e_{t+1}} \in \mathcal{E}} \sum\limits_{{{\rm{s}}_{t+1} } \in \mathcal{S}} q_\sigma(e_{t+1})q_\sigma({s_{t+1}}|{e_{t+1}})$. By using recursion for the last term in the third equation, Lemma 1 can be derived.

\end{proof}

\section{Proof of Theorem 1}
\begin{proof}
Based on Lemma 1, there is
\begin{equation}
\label{EquationPD}
\begin{split}
& J_{t,\sigma}(e_t) - J_{t,\upsilon}(e_t) = \\
& \sum\limits_{\tau  = t}^{t + T_w - 1}  \sum\limits_{{e_\tau} \in \mathcal{E}} \sum\limits_{{{\rm{s}}_\tau } \in \mathcal{S}} q_\sigma(e_\tau)q_\sigma({s_\tau }|{e_\tau }) [{r_{\tau ,\sigma }} - {r_{\tau ,\upsilon }} + ({p_{\tau ,\sigma }} - {p_{\tau ,\upsilon }})J_{\tau  + 1,\upsilon } ] \\
& + \beta \sum\limits_{\tau  = t}^{t + T_w - 1}  \sum\limits_{{e_\tau} \in \mathcal{E}} \sum\limits_{{{\rm{s}}_\tau } \in \mathcal{S}} [{{({c_{\tau ,\upsilon }} - {{\bar J}_\upsilon })}^2} - {{({c_{\tau ,\sigma }} - {{\bar J}_\sigma })}^2}]
\end{split}
\end{equation}
Let $F = \sum\limits_{\tau  = t}^{t + T_w - 1}  \sum\limits_{{e_\tau} \in \mathcal{E}} \sum\limits_{{{\rm{s}}_\tau } \in \mathcal{S}} q_\sigma(e_\tau)q_\sigma({s_\tau }|{e_\tau }) [{{({c_{\tau ,\upsilon }} - {{\bar J}_\upsilon })}^2} - {{({c_{\tau ,\sigma }} - {{\bar J}_\sigma })}^2}]$. As $\sum\limits_{\tau  = t}^{t + T_w - 1}  \sum\limits_{{e_\tau} \in \mathcal{E}} \sum\limits_{{{\rm{s}}_\tau } \in \mathcal{S}} q_\sigma(e_\tau)q_\sigma({s_\tau }|{e_\tau }) = T_w$ and $\sum\limits_{\tau  = t}^{t + T_w - 1}  \sum\limits_{{e_\tau} \in \mathcal{E}} \sum\limits_{{{\rm{s}}_\tau } \in \mathcal{S}} q_\sigma(e_\tau)q_\sigma({s_\tau }|{e_\tau }) {c_{\tau ,\sigma }} = T_w {\bar J}_\sigma$, there is
\begin{equation}
\label{EquationTemp1}
\begin{split}
& \sum\limits_{\tau  = t}^{t + T_w - 1} \sum\limits_{{e_\tau} \in \mathcal{E}} \sum\limits_{{{\rm{s}}_\tau } \in \mathcal{S}} q_\sigma(e_\tau)q_\sigma({s_\tau }|{e_\tau }) [ - {\bar J}_{\sigma}^2 + 2 c_{\tau ,\sigma } {\bar J}_\sigma] \\
& = {\bar J}_\sigma \sum\limits_{\tau  = t}^{t + T_w - 1} \sum\limits_{{e_\tau} \in \mathcal{E}} \sum\limits_{{{\rm{s}}_\tau } \in \mathcal{S}} q_\sigma(e_\tau)q_\sigma({s_\tau }|{e_\tau }) [ - {\bar J}_{\sigma} + 2 c_{\tau ,\sigma }] \\
& = {\bar J}_\sigma [-T_w{\bar J}_\sigma + 2T_w {\bar J}_\sigma] = T_w{\bar J}_\sigma^2
\end{split}
\end{equation}
Substituting (\ref{EquationTemp1}) into $F$, we can derive
\begin{equation}
\label{EquationTemp2}
\begin{split}
& F = \sum\limits_{\tau  = t}^{t + T_w - 1}  \sum\limits_{{e_\tau} \in \mathcal{E}} \sum\limits_{{{\rm{s}}_\tau } \in \mathcal{S}} q_\sigma(e_\tau)q_\sigma({s_\tau }|{e_\tau }) \\
& [c_{\tau ,\upsilon }^2 + {\bar J}_\upsilon^2 - 2 c_{\tau ,\upsilon } {\bar J}_\upsilon - c_{\tau ,\sigma }^2 - {\bar J}_{\sigma}^2 + 2 c_{\tau ,\sigma } {\bar J}_\sigma] \\
& = \sum\limits_{\tau  = t}^{t + T_w - 1}  \sum\limits_{{e_\tau} \in \mathcal{E}} \sum\limits_{{{\rm{s}}_\tau } \in \mathcal{S}} q_\sigma(e_\tau)q_\sigma({s_\tau }|{e_\tau }) \\
& [c_{\tau ,\upsilon }^2 + {\bar J}_\upsilon^2 - 2 c_{\tau ,\upsilon } {\bar J}_\upsilon - c_{\tau ,\sigma }^2 ] + T_w{\bar J}_\sigma^2 \\
& = \sum\limits_{\tau  = t}^{t + T_w - 1}  \sum\limits_{{e_\tau} \in \mathcal{E}} \sum\limits_{{{\rm{s}}_\tau } \in \mathcal{S}} q_\sigma(e_\tau)q_\sigma({s_\tau }|{e_\tau }) [c_{\tau ,\upsilon }^2 + {\bar J}_\upsilon^2 \\
& - 2 c_{\tau ,\upsilon } {\bar J}_\upsilon - c_{\tau ,\sigma }^2 ] + T_w({\bar J}_\sigma - {\bar J}_\upsilon)^2 + 2T_w{\bar J}_\sigma {\bar J}_\upsilon - T_w{\bar J}_\upsilon^2\\
\end{split}
\end{equation}
For (\ref{EquationTemp2}), there is
\begin{equation}
\label{EquationTemp3}
\begin{split}
& -2 \sum\limits_{\tau  = t}^{t + T_w - 1}  \sum\limits_{{e_\tau} \in \mathcal{E}} \sum\limits_{{{\rm{s}}_\tau } \in \mathcal{S}} q_\sigma(e_\tau)q_\sigma({s_\tau }|{e_\tau }) c_{\tau ,\upsilon } {\bar J}_\upsilon +  2T_w{\bar J}_\sigma {\bar J}_\upsilon \\
& = 2 {\bar J}_\upsilon [-\sum\limits_{\tau  = t}^{t + T_w - 1}  \sum\limits_{{e_\tau} \in \mathcal{E}} \sum\limits_{{{\rm{s}}_\tau } \in \mathcal{S}} q_\sigma(e_\tau)q_\sigma({s_\tau }|{e_\tau }) c_{\tau ,\upsilon } \\
& + \sum\limits_{\tau  = t}^{t + T_w - 1}  \sum\limits_{{e_\tau} \in \mathcal{E}} \sum\limits_{{{\rm{s}}_\tau } \in \mathcal{S}} q_\sigma(e_\tau)q_\sigma({s_\tau }|{e_\tau }) {c_{\tau ,\sigma }}] \\
& = 2 {\bar J}_\upsilon \sum\limits_{\tau  = t}^{t + T_w - 1}  \sum\limits_{{e_\tau} \in \mathcal{E}} \sum\limits_{{{\rm{s}}_\tau } \in \mathcal{S}} q_\sigma(e_\tau)q_\sigma({s_\tau }|{e_\tau }) ({c_{\tau ,\sigma}} - {c_{\tau ,\upsilon}}) \\
\end{split}
\end{equation}
Substituting (\ref{EquationTemp3}) into (\ref{EquationTemp2}) and using $\sum\limits_{\tau  = t}^{t + T_w - 1}  \sum\limits_{{e_\tau} \in \mathcal{E}} \sum\limits_{{{\rm{s}}_\tau } \in \mathcal{S}} q_\sigma(e_\tau)q_\sigma({s_\tau }|{e_\tau }) {\bar J}_\upsilon^2 = T_w {\bar J}_\upsilon^2$, there is
\begin{equation}
\label{EquationTemp4}
\begin{split}
& F = T_w({\bar J}_\sigma - {\bar J}_\upsilon)^2 + \sum\limits_{\tau  = t}^{t + T_w - 1}  \sum\limits_{{e_\tau} \in \mathcal{E}} \sum\limits_{{{\rm{s}}_\tau } \in \mathcal{S}} q_\sigma(e_\tau)q_\sigma({s_\tau }|{e_\tau }) \\
& [c_{\tau ,\upsilon }^2 - c_{\tau ,\sigma }^2 + 2 {\bar J}_\upsilon ({c_{\tau ,\sigma}} - {c_{\tau ,\upsilon}}) ] \\
\end{split}
\end{equation}
Substituting (\ref{EquationTemp4}) into (\ref{EquationPD}), we can derive
\begin{equation}
\label{EquationPDNew}
\begin{split}
& J_{t,\sigma}(e_t) - J_{t,\upsilon}(e_t) =  \sum\limits_{\tau  = t}^{t + T_w - 1}  \sum\limits_{{e_\tau} \in \mathcal{E}} \sum\limits_{{{\rm{s}}_\tau } \in \mathcal{S}} q_\sigma(e_\tau)q_\sigma({s_\tau }|{e_\tau }) \\
& \{ [{r_{\tau ,\sigma }} + {p_{\tau ,\sigma }}J_{\tau  + 1,\upsilon } - \beta ({c_{\tau ,\sigma }^2} - 2{c_{\tau ,\sigma }} {\bar J}_\upsilon )] \\
& - [{r_{\tau ,\upsilon }} + {p_{\tau ,\upsilon }}J_{\tau  + 1,\upsilon } - \beta ({c_{\tau ,\upsilon }^2} - 2{c_{\tau ,\upsilon }} {\bar J}_\upsilon )] \} + \beta T_w({\bar J}_\sigma - {\bar J}_\upsilon)^2 \\
& = \sum\limits_{\tau  = t}^{t + T_w - 1}  \sum\limits_{{e_\tau} \in \mathcal{E}} \sum\limits_{{{\rm{s}}_\tau } \in \mathcal{S}} q_\sigma(e_\tau)q_\sigma({s_\tau }|{e_\tau }) \\
&\{ [{r_{\tau ,\sigma }} + {p_{\tau ,\sigma }}J_{\tau  + 1,\upsilon } - \beta ({c_{\tau ,\sigma }} - {\bar J}_\upsilon )^2] \\
& - [{r_{\tau ,\upsilon }} + {p_{\tau ,\upsilon }}J_{\tau  + 1,\upsilon } - \beta ({c_{\tau ,\upsilon }} - {\bar J}_\upsilon )^2] \} + \beta T_w({\bar J}_\sigma - {\bar J}_\upsilon)^2
\end{split}
\end{equation}

From (\ref{EquationPDNew}), it can be seen that if the condition in theorem 1 is satisfied, there is $J_{t,\sigma}(e_t) \geq J_{t,\upsilon}(e_t)$. Furthermore, if there exists $e_{\tau^{'}}, \tau^{'} = t,t+1,...,t+T_w-1$ which the inequality strictly holds, there is $J_{t,\sigma}(e_t) > J_{t,\upsilon}(e_t)$.
\end{proof}

\section{Proof of Theorem 2}
\begin{proof}
We prove it by using contradiction. Suppose equation (\ref{EquationOptimalPolicy}) does not hold for optimal policy $\sigma^*$. As the initial distribution $q_\upsilon({s_t }|{e_t })$ is independent with the policies, we can at least find a action $a_{t}^{'}$ which has
\begin{equation}
\begin{split}
& \sum\limits_{{{\rm{s}}_t } \in \mathcal{S}} q_\sigma({s_t }|{e_t}) [r_{t}^{'} - \beta {{({c_{t}^{'}} - {{\bar J}_{\sigma^*} })}^2} + {p_{t}^{'}}{J_{t + 1,\sigma^* }}] \\
& > \sum\limits_{{{\rm{s}}_t } \in \mathcal{S}} q_\sigma({s_t }|{e_t }) [{r_{t ,\sigma^*}} - \beta {{({c_{t ,\sigma^*}} - {{\bar J}_\sigma^* })}^2} + {p_{t ,\sigma^* }}{J_{t  + 1,\sigma^* }}]
\end{split}
\end{equation}
where ${c_{t}^{'}}$, $r_{t}^{'}$ and ${p_{t}^{'}}$ are the incurred values corresponding to $a_{t}^{'}$. Therefore, a new policy $\upsilon$ can be constructed by the following mechanism,
\begin{equation}
\upsilon(e_t) = a_{t}^{'}, \upsilon(e_\tau) = \sigma^*(e_\tau), \forall \tau = t+1,...,t+T_w-1.
\end{equation}
Based on (\ref{EquationPDNew}), we can derive $J_{t,\upsilon}(e_t) - J_{t,\sigma^*}(e_t) > \beta T({\bar J}_\sigma - {\bar J}_\upsilon)^2 \geq 0 $. This means $\sigma^*$ is not the optimal event-based scheduling policy which is contradictory. Thus, based on the above discussions, Theorem 2 is proven.

\end{proof}

%% use section* for acknowledgment
%\section*{Acknowledgment}

\ifCLASSOPTIONcaptionsoff
  \newpage
\fi

\bibliographystyle{IEEEtran}
\bibliography{IEEEabrv,myref}

\end{document}